\documentclass[a4paper,11pt]{article}

\pdfoutput=1 

\usepackage{jheppub} 

\usepackage{graphicx}
\usepackage{subfigure}
\usepackage{multirow}
\usepackage{ulem}
\usepackage{color}
\definecolor{My_red}        {cmyk}{0.00,1.00,1.00,0.20}

\usepackage{slashed}

\newcommand{\met}{$\slashed{E}_T$~}

\newcommand{\bmat}{\left(\begin{array}}
\newcommand{\emat}{\end{array}\right)}
\newcommand{\beq}{\begin{equation}}
\newcommand{\eeq}{\end{equation}}

\newcommand{\VEV}[1]{\langle  #1 \rangle}

\def\bwt{\begin{widetext}}
\def\ewt{\end{widetext}}
\def\be{\begin{equation}}
\def\ee{\end{equation}}
\def\bea{\begin{eqnarray}}
\def\eea{\end{eqnarray}}
\def\bean{\begin{eqnarray*}}
\def\eean{\end{eqnarray*}}
\def\bary{\begin{array}}
\def\eary{\end{array}}
\def\bit{\begin{itemize}}
\def\eit{\end{itemize}}

\def\su5u1{SU(5) \times U(1)}
\def\fsu5u1{SU(5) \times U(1)'}
\def\so10{SO(10)}
\def\sq20{SO(10) \times SO(10)}

\def\nn{\nonumber}

\def\bwt{\begin{widetext}}
\def\ewt{\end{widetext}}
\def\be{\begin{equation}}
\def\ee{\end{equation}}
\def\bea{\begin{eqnarray}}
\def\eea{\end{eqnarray}}
\def\bean{\begin{eqnarray*}}
\def\eean{\end{eqnarray*}}
\def\bary{\begin{array}}
\def\eary{\end{array}}
\def\bit{\begin{itemize}}
\def\eit{\end{itemize}}

\def\su5u1{SU(5) \times U(1)}
\def\fsu5u1{SU(5) \times U(1)'}
\def\so10{SO(10)}
\def\sq20{SO(10) \times SO(10)}

\title{Diphoton Excess in Consistent Supersymmetric ${\boldsymbol{SU(5)}}$ Models with Vector-like Particles}

\author[a]{Bhaskar Dutta}
\author[a]{Yu Gao }
\author[a,1]{Tathagata Ghosh}
\author[b,2]{Ilia Gogoladze, \note{On  leave of absence from:
Andronikashvili Institute of Physics, 0177 Tbilisi, Georgia.}}
\author[c,d]{Tianjun Li }
\author[b]{Qaisar Shafi }
\author[e]{Joel W. Walker,\note{Corresponding author.}}

\affiliation[a]{
Mitchell Institute for Fundamental Physics and Astronomy,
Department of Physics and Astronomy, Texas A\&M University, College Station, TX 77843-4242, USA }

\affiliation[b]{
Bartol Research Institute, Department of Physics and Astronomy,
University of Delaware, Newark, DE 19716, USA }

\affiliation[c]{State Key Laboratory of Theoretical Physics and
Kavli Institute for Theoretical Physics China (KITPC),
Institute of Theoretical Physics, Chinese Academy of Sciences,
Beijing 100190, P. R. China}

\affiliation[d]{
School of Physical Electronics, University of Electronic Science and Technology of China,
Chengdu 610054, P. R. China
}

\affiliation[e]{
Department of Physics, Sam Houston State University, Huntsville, TX 77341, USA}

\emailAdd{dutta@physics.tamu.edu}
\emailAdd{yugao@physics.tamu.edu}
\emailAdd{ghoshtatha@physics.tamu.edu}
\emailAdd{ilia@bartol.udel.edu}
\emailAdd{tli@itp.ac.cn}
\emailAdd{shafi@bartol.udel.edu}
\emailAdd{jwalker@shsu.edu}

\abstract{We consider the diphoton resonance at the 13 TeV LHC in the context of $SU(5)$ grand unification.
A leading candidate to explain this resonance is a standard model singlet
scalar decaying to a pair of photon by means of vector-like fermionic loops.
We demonstrate the effect of the vector-like multiplets $(5,\overline{5})$ and $(10,\overline{10})$ on the
evolution of the gauge couplings and perturbatively evaluate the weak scale values of
the new couplings and masses run down from the unification scale.
We use these masses and couplings to explain the diphoton resonance after considering the new dijet constraints.
We show how to accommodate a potentially large decay width of the resonance particle or particles, and introduce
a method for quantifying the statistical preference of data for various resonance width and mass gap scenarios.
In addition, we consider new couplings relating various components of $(5,\overline{5})$ and
$(10,\overline{10})$ in the context of the orbifold GUTs, where the resonance scalar can be a part of the new
vector-like lepton doublets.  We also calculate the Higgs mass and proton decay rate $p\rightarrow e^{+}\pi^0$
in the context of $SU(5)$ grand unification, including effects of the new vector-like multiplets.
Although the specific excess described this work has not carried over into the newly released 2016
data, the analysis presented remains more broadly applicable to the general topic of scattering through a resonance,
especially in the context of diphoton final states with loops of vector-like multiplets.
}

\begin{document}

\begin{flushright}
MI-TH-1601
\end{flushright}




\maketitle

\section{Introduction}

Supersymmetry (SUSY) provides an elegant solution to the gauge hierarchy problem in the Standard Model (SM).
In addition it has many appealing features.
In Supersymmetric SMs (SSMs) with R-parity, we can realize gauge coupling unification,
have the Lightest Supersymmetric Particle (LSP), namely the neutralino, as a dark matter candidate,
radiatively break the electroweak gauge symmetry, etc.
In particular, gauge coupling unification strongly supports the Grand Unified Theories (GUTs),
and supersymmetry is thus a bridge between the low energy phenomenology and high-energy fundamental physics.

It is well known that a SM-like Higgs boson with mass $m_H$ around 125 GeV was discovered
during the first run of the LHC~\cite{ATLAS, CMS}. In the Minimal SSM (MSSM), to realize
such a Higgs boson mass it is necessary to have either multi-TeV top squarks with small mixing or TeV-scale
top squarks with large mixing, which might increase the fine-tuning or induce
$SU(3)\times U(1)_{\rm EM}$ gauge symmetry breaking~\cite{breakingvac}, respectively.
On the other hand, it has long been understood that one can extend the matter sector of
the MSSM and still preserve the elegant result of gauge coupling
unification if the new matter fields form complete multiplets
of $SU(5)$. To automatically cancel the gauge anomalies, we assume here that such supermultiplets are vector-like.
In fact, complete light GUT multiplets of this variety are not unexpected.  Within
string theory one often finds light multiplets in the spectrum~\cite{Dienes:1996du}, and
even within the GUT framework itself one can encounter extra complete
multiplets lying at the TeV scale~\cite{Babu:1996zv}.
Similarly to the top quark contribution, the Higgs boson mass can be lifted by the Yukawa couplings
between these vector-like particles and the Higgs fields. Consequently, the SM gauge couplings
will become stronger at the GUT scale, in which case the proton lifetime will be reduced,
thus coming within the reach of proton decay experiments such as Hyper-Kamiokande.

Recently, both the ATLAS~\cite{bib:ATLAS_diphoton} and CMS~\cite{bib:CMS_diphoton} collaborations have reported an
excess of events in the diphoton channel with invariant mass of about 750 GeV at the 13 TeV LHC.
With an integrated luminosity of 3.2 ${\rm fb}^{-1}$, the ATLAS collaboration has observed a local $3.6 \, \sigma$
excess at a diphoton invariant mass of around 747~GeV, assuming
a narrow width resonance. For a wider width resonance, the signal significance
increases to $3.9 \, \sigma$ with a preferred width of about 45~GeV. They also found about $1.9 \, \sigma$ excess when they reanalyze their 8 TeV data~\cite{bib:ATLAS_moriond}.
With an integrated luminosity of 2.6 ${\rm fb}^{-1}$, the CMS collaboration has also observed a diphoton excess
with a local significance of $2.6 \, \sigma$ at invariant mass of around 760 GeV. Assuming
a decay width of around 45~GeV, the significance reduces to $2 \, \sigma$ in this case.
The corresponding excesses in the cross section can be roughly estimated as
$\sigma_{pp\to \gamma \gamma}^{13~ {\rm TeV}} \sim 3-13~{\rm fb}$~~\cite{bib:ATLAS_diphoton, bib:CMS_diphoton}.
Interestingly, the CMS collaboration did likewise search for diphoton resonances~\cite{Khachatryan:2015qba}
at $\sqrt{s} =$ 8 TeV and observed a slight excess $\sim$ $2 \, \sigma$ at an invariant mass of about 750 GeV.
Thus, the present ATLAS and CMS
results at $\sqrt{s} =$~13~TeV are indeed consistent with those at $\sqrt{s} =$~8 TeV LHC in the diphoton
channel. The dijet constraints from LHC may strongly constrain any interpretation of this resonance.
The CMS collaboration has recently reported a new analysis on dijet final state resonances~\cite{CMS:2015neg}.
We investigated the compatibility of our explanation of the resonance against this result.

A straightforward approach to explaining the diphoton excess is the introduction of a SM singlet $S$ with
mass of 750~GeV and accompanying multiplets of vector-like particles.
With vector-like particles in the loops, the singlet $S$ can be produced via gluon fusion (similar to SM Higgs cases previously considered in Refs.~\cite{ggH,Joglekar}),
and can likewise decay into a diphoton pair. The diphoton excess, in a non-supersymmetric context, was previously addressed by some of the authors
using vector-like particles, as motivated by solutions to the gauge unification, neutrino mass,
and electroweak vacuum stability problems~\cite{Dutta:2015wqh}.
This approach can be naturally embedded in the SSMs with vector-like particles.
In this paper, we study gauge coupling unification, calculate the Higgs boson mass, and
estimate the diphoton event rate.
After careful study, we find that the there is some delicacy to the assignment of
multiplicities and couplings for vector-like multiplets in a manner which
explains the diphoton resonance while still achieving the gauge coupling unification.
A recent analysis~\cite{Falkowski} of CMS and ATLAS data and a fit to the combined run-I
and run-II data indicates that the resonance at 750 GeV can be accommodated by
$\sigma_{\gamma \gamma} \sim 0.7 - 16$~fb for $\Gamma_S \sim 5-100$ GeV at $2 \sigma$ level.
We explore the capacity of $(10,\overline{10})$ and $(5,\overline{5})$ multiplets to
explain the resonance. We additionally demonstrate how to accommodate a larger resonance
width in the context of this unified scenario by introducing decays of the scalar to soft leptons
with very little missing energy, which is allowed by current experimental data.
We point out that the proton lifetime estimates lie within reach of the future
Hyper-Kamiokande experiment.

We also show that the neutral component of the vector-like lepton doublets can likewise be utilized to
explain the excess. In this case these doublets are R-parity even, which moreover will not induce proton decay
problems.  Such scenarios can be realized in the orbifold GUTs~\cite{kawa, GAFF, LHYN}
and F-theory GUTs~\cite{Donagi:2008ca, Beasley:2008dc, Beasley:2008kw, Donagi:2008kj}
(see Ref.~\cite{Li:2010hi} and references therein). The requirement of two new doublets in this scenario allows for the possibility of two adjacent resonances, which can be useful to generate the appearance of a large effective width.


This excess, although still statistically short of discovery, has drawn immense attention from the
particle physics community, resulting in diverse explanations ranging from axions and extended Higgs sectors to dark
matter~\cite{Dutta:2015wqh,Falkowski,diphoton-first,diphoton-rest,diphoton_request,Hall:2015xds,Patel:2015ulo,Ding:2015rxx,Allanach:2015ixl}.
In two recent papers~\cite{Hall:2015xds,Patel:2015ulo},
the diphoton excess has been addressed in the context of $SU(5)$ grand unification.
Vector-like multiplet masses in the context of a SUSY $SU(5)$ GUT are studied in Ref.~\cite{Hall:2015xds}, 
while Ref.~\cite{Patel:2015ulo} considers enhancement of the cross-section and diphoton branching ratio
of a singlet type resonance due to the presence of 24 dimensional adjoint fermions at the sub-TeV scale.
$SU(5)$ grand unification with vector-like fermion multiplets was also considered in the past~\cite{Joglekar},
to explain the SM $H \rightarrow \gamma \gamma$ excess previously observed at the LHC. In our paper, however,
we have included the new dijet constraint in order to estimate the number of vector-like multiplets necessary
to address the diphoton excess, which we then utilize to calculate the Higgs mass.
We additionally consider a new interaction involving
the neutral scalar from vector-like doublets to explain the excess.
We also show new ways to handle the larger resonance width and associated final states.

The paper is organized as follows.
In section II, we discuss the GUT models with vector-like particles and projections for proton decay.
In section III, we discuss the impact on Higgs Boson mass arising from vector-like particles.
In section IV, we discuss the neutral component of vector-like doublets as a resonance candidate, and
in section V, we discuss the diphoton resonance, and introduce a statistical test for quantifying the decay width
preferred by binned data in the context of a dual resonance.
We conclude in section VI.

\section{GUT Models with Vector-Like Particles and Proton Decay}

It is well known that matter fields will contribute at one loop to the
CP-even Higgs mass if there are direct couplings between them and
the Higgs fields.  We will elaborate upon these additional contributions in the next section.
On the other hand, there are constraints on the couplings and masses of new matter
fields if they are involved in chiral symmetry breaking interactions.
The most important constraints are the $S$ and $T$ parameters,
which limit the number of extra {\it chiral}\/ generations. Consistent with these
constraints, one should add new matter fields which are predominantly vector-like.

In the limit where the vector-like mass is much heavier than the
chiral mass terms arising from Yukawa couplings to the
Higgs doublets, the contribution to the $T$ parameter from a
single chiral fermion is approximately~\cite{Lavoura:1992np}:
\begin{equation}
\delta T=\frac{ N (\kappa v)^2}{10 \pi \sin^2\theta_W m^2_W}\left[ \left(
\frac{\kappa v}{M_V}\right)^2  +O\left( \frac{\kappa
v}{M_V}\right)^4\right], \label{tpar}
\end{equation}
where $\kappa$ is the new chiral Yukawa coupling, $v$ is the
Vacuum Expectation Value (VEV) of the corresponding Higgs field,
and $N$ counts the additional
number of $SU(2)$ doublets. For instance, $N=3$ if $(10, \overline {10})$
is considered at low scale, while  $N=1$ for the $(5,\overline 5)$
case. It is known that from precision electroweak data $T\leq 0.2$ at 95\% CL for
$m_h=125$ GeV~\cite{Agashe:2014kda}. We
will take $\delta T<0.2$ as a realistic bound in our analysis.
We then see from Eq.~(\ref{tpar}) that with $M_V$ around 1~TeV, the
Yukawa coupling $\kappa$ can be $O(1)$.

A strong constraint on the nature of new vector-like particles arises
from the perturbativity and unification conditions.
One finds that the following combinations of low-energy (TeV-scale)
vector-like particles may be introduced safely:
{\it (i)} up to 4 pairs of $(5,\overline{5})$'s,
or {\it (ii)} one pair of $(10,\overline{10})$,
or {\it (iii)} one pair each of $(5,\overline{5})$ and $(10,\overline{10})$.
The last option also neatly fits into the $(16+\overline{16})$
representation of $SO(10)$ if an additional pair of singlets are added.
We will thus refer to case {\it (iii)} as $(16+\overline{16})$.

\begin{figure}[t]
\centering \includegraphics[angle=0, width=12cm]{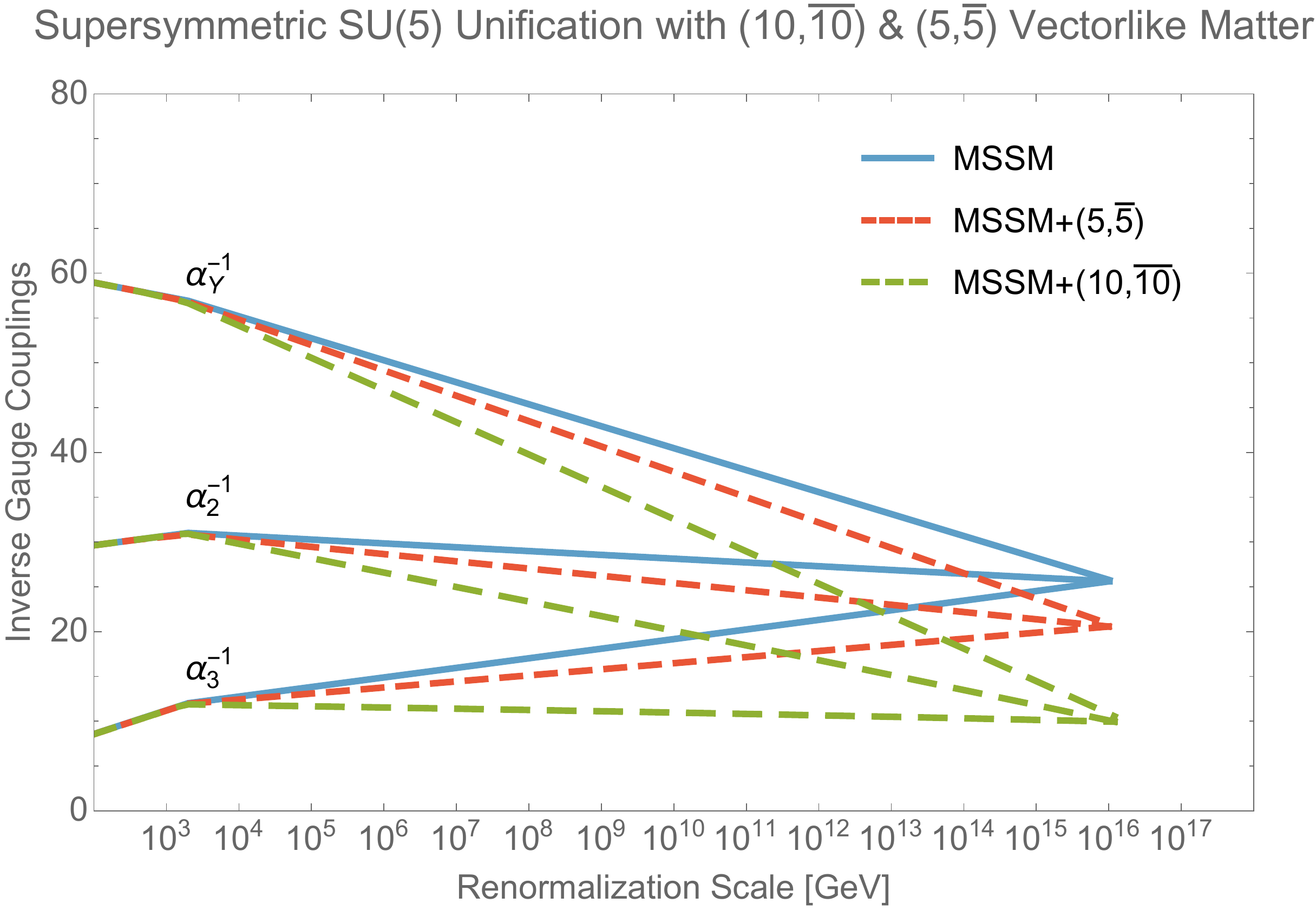}
\vspace{-0.3cm} \caption{ Gauge coupling evolution with the effective
SUSY breaking scale $M_{\text SUSY}= 2$ TeV and tan$\protect\beta = 10$.
Solid lines correspond to the MSSM. Short dashed lines
correspond to the MSSM$+(5,\overline{5})$. Long dashed lines are for
MSSM plus $(10,\overline{10})$, which is essentially the same as MSSM
plus $3\times(5,\overline{5})$.
Vector-like masses are set at $M_V^Q=1$~TeV and $M_V^L=400$~GeV.} \label{gu}
\end{figure}

We illustrate in Figure~\ref{gu} how gauge coupling running is modified by
introducing different sets of vector-like particles.
In particular, examples of the gauge coupling evolution are plotted
for the case of MSSM by itself and MSSM plus the complete $SU(5)$ multiplets $(10 + \overline{10})$ and $(5 + \overline{5})$.
The GUT-scale $M_{\rm GUT}$ and unified coupling $\alpha_{\rm GUT}$ applicable to
each of the itemized scenarios are further presented numerically in Table~\ref{tab:proton}.
RGEs are run at two loops in the gauge sector,
with feedback from the one loop MSSM Yukawa couplings.
For uniformity, all entries in Table~\ref{tab:proton}, as well as Figure~\ref{gu},
are computed for colored and non-colored vector-like masses of $M_V^Q=1$~TeV and $M_V^L=400$~GeV,
with all sparticles (including the scalar vector-like partners) at $M_{\rm SUSY}=2$~TeV,
and $\tan \beta = 10$.
The residual gap $\Delta_{Y2} \equiv \vert g_Y - g_2 \vert \div (g_Y + g_2)$ between the
hypercharge and $SU(2)_{\rm L}$ couplings at the scale where the perturbative
unification $\alpha_3 = \alpha_2$ occurs is less well controlled in the $(16+\overline{16})$
scenario, but this may be mitigated substantially by elevating the SUSY scale
into the several TeV range.

\begin{table}[h]
\begin{center}
\bgroup
\def\arraystretch{1.5}%
\begin{tabular}{|c|c|c|c|c|c|c|}
\hline \hline
$N(10,\overline{10})$ & $N(5,\overline{5})$ & $M_{\rm GUT}$ & $\alpha_{\rm GUT}$ &
$\Delta_{Y2}$ & $A_R^{SD}$ & $\tau_{\rm p} (e^+\pi^0)$ \\ \hline
 0 & 0 & 1.1 & 0.039 & 0.001 & 2.0 & 14 \\ \hline
 0 & 1 & 1.1 & 0.049 & 0.003 & 2.4 & 6.9 \\ \hline
 0 & 2 & 1.2 & 0.065 & 0.004 & 3.1 & 3.2 \\ \hline
 0 & 3 & 1.4 & 0.10 & 0.007 & 4.3 & 1.5 \\ \hline
 0 & 4 & 2.7 & 0.30 & 0.007 & 6.7 & 0.93 \\ \hline
 1 & 0 & 1.2 & 0.10 & 0.016 & 4.3 & 0.76 \\ \hline
 1 & 1 & 1.6 & 0.30 & 0.070 & 6.8 & 0.10 \\ \hline \hline
\end{tabular}
\egroup
\caption{
Unification parameters and proton lifetime projections for various configurations
of $SU(5)$ vector-like supermultiplets taken in addition to the field content of the MSSM.
The scale $M_{\rm GUT}$ at which $\alpha_3 = \alpha_2 = \alpha_{\rm GUT}$ is given in units of $10^{16}$~[GeV].
The ratio $\Delta_{Y2}$ represents the fractional separation of $\alpha_Y$ and $\alpha_2$ at the GUT scale.
The dimensionless factors $A_R^{SD}$ reflect short-distance renormalization of the anomalous dimension
associated with relevant baryon-number violating operators.  The proton lifetime $\tau_{\rm p}$
in the dimension-six $e^+\pi^0$ channel is projected in units of $10^{34}$~[y].
All entries are computed for vector-like masses $M_V^Q=1$~TeV and $M_V^L=400$~GeV,
with all sparticles at $M_{\rm SUSY}=2$~TeV, and $\tan \beta = 10$.
}
\label{tab:proton}
\end{center}
\end{table}

We next consider the rate of proton decay $p \to e^+ \pi^0$ via dimension-6
operators from heavy gauge boson exchange,
in keeping with the prescription of Ref.~\cite{Hisano:2000dg}.
\begin{equation}
\tau_p(e^+\pi^0) \simeq 1.0\times 10^{34} \times
\left(\frac{2.5}{A_R}\right)^2 \times
\left(\frac{0.04}{\alpha_{\rm GUT}}\right)^2 \times
\left(\frac{M_{\rm GUT}}{1.0\times 10^{16}~{\rm GeV}}\right)^4
~{\rm years}
\label{eq:proton}
\end{equation}
The lifetime scales as a fourth power of the unification scale $M_{\rm GUT}$, as an inverse-squared power
of the unified coupling $\alpha_{\rm GUT}$, as an inverse-squared power of the hadronic matrix
element $\alpha_{\rm H}$~\cite{Aoki:1999tw}, and as an inverse-squared power of the dimensionless 1-loop renormalization
factor $A_R \equiv A_R^{SD}A_R^{LD}$ associated with anomalous dimension of the relevant baryon-number
violating operators.  The long-distance factor $A_R^{LD}$ takes
a universal value of approximately 1.2, while the short-distance factor $A_R^{SD}$ is highly
dependent upon the ultra-violet field content, generally increasing with the addition of new vector-like
supermultiplets.  For the special cases $(10,\overline{10})$ and $3\times(5,\overline{5})$,
where the 1-loop beta-function coefficient of the strong coupling vanishes, a limit for
the continuous value of $A_R^{SD}$ may be smoothly numerically extrapolated.
The central projected proton lifetime for each of the itemized scenarios is presented in Table~\ref{tab:proton}.
Current limits on the considered $e^+\pi^0$
decay mode are around $1.7\times10^{34}$~years~\cite{Kearns}.  Uncertainties in the hadronic matrix
element~\cite{Aoki:2013yxa}, the finite-order renormalization group analysis, the low-energy boundary
values, and unknown high-energy threshold effects, coupled with the large powers apparent
in Eq.~(\ref{eq:proton}) lead to substantial uncertainties in the projected rate, often estimated
to exceed an order of magnitude~\cite{Kearns}.  It would seem then that all scenarios considered
in Table~~\ref{tab:proton}, with the possible exception of the $(10,\overline{10}+5,\overline{5})$
case, are generally consistent with current bounds.  Moreover, several of these scenarios
point to a high likelihood of a signal at next-generation experiments such as Hyper-Kamiokande~\cite{Abe:2011ts}.
We assume that the potentially dangerous dimension-five higgsino-mediated proton decay has been
appropriately suppressed.  We remark that this operator is naturally suppressed in the
flipped $SU(5)$ GUTs, and the $e^+\pi^0$ lifetime is simultaneously extended by a factor of about
five due to absence of $\overline{10}10\overline{10}10$ type contributions~\cite{Ellis:2002vk}, although
we do not consider those scenarios further here.

\section{The Higgs Boson Mass and Vector-Like Particles}

\subsection{{MSSM }+ $\mathbf{(10,\overline{10})}$}

As previously described, if there is direct coupling among new matter fields and
the MSSM Higgs field, the new matter fields will contribute at one-loop level to the
CP-even Higgs mass. Here, we consider in detail the case when new vector-like particles
fill up $(10,\overline{10})$ dimensional representation of $SU(5)$.
The representation $(10,\overline{10})$ of $SU(5)$
decomposes under the MSSM gauge symmetry as follows:%
\begin{eqnarray}
  10 &=&Q_{10}\left( 3,2,\frac{1}{6}\right)
  + U_{10}\left(\overline{3},1,-\frac{2}{3}\right)
+E_{10}\left( 1,1,1\right) ~,~\nonumber \\
  \overline{10} &=& \overline{Q}_{10}\left( \overline{3},2,-\frac{1}{6}\right)
   +\overline{U}_{10}\left( 3,1,\frac{2}{3}\right) +\overline{E}_{10}\left( 1,1,-1\right).
\end{eqnarray}

In case when we have new vector-like particles from $(10,\overline{10})$ multiplet, the new couplings
$10\cdot 10\cdot H_u$ and ${\overline{10}}\cdot {\overline{10}}\cdot H_d$
are allowed, analogous to the top--quark Yukawa
couplings, but involving the charge $2/3$ ($-2/3$) quark from the $10$-plet (${\overline{10}}$-plet).
Note that we employ the $SU(5)$ notation here for simplicity, with the
understanding that $H_u$ and $H_d$ are not complete multiplets of $SU(5)$.
Here we assume that the model also contains the SM gauge singlet $S$ field.
A new coupling ($S 10\overline{10}$) is then allowed.

The part of the superpotential describing interaction among $(10,\overline{10})$, $S$
and the MSSM Higgs fields has the following form: %
\begin{eqnarray}
W&=&\kappa _{10}Q_{10}\, U_{10}\, H_{u}+\kappa _{10}^{\prime }\overline{Q}_{10}
\, \overline{U}_{10}\, H_{d}+ \lambda^Q_{10} S\, \overline{Q}_{10} \,Q_{10} + \lambda^U_{10} S \,\overline{U}
_{10}\, U_{10} + \lambda^E_{10} S\, \overline{E}_{10}\, E_{10}
 \nonumber \\
&+& \lambda S \,H_{u}\,  H_{d} + m_S S^2 + M_{V}\left( \overline{Q}_{10}\,Q_{10}+\overline{U}
_{10}\,U_{10}+\overline{E}_{10}\,E_{10}\right), \label{nn1}
\end{eqnarray}%
where we have taken a common vector-like mass at the
GUT scale $M_{\rm GUT}$ for simplicity.
Thus, the up quark-like pieces of the $10$ and
$\overline{10}$ acquire Dirac {\it and}\/ vector-like masses, while
the $E_{10}$ lepton-like pieces receive only vector-like masses. In the case that 
$\kappa _{10}\gg \kappa _{10}^{\prime }$, then the contribution
coming from the coupling $ \kappa _{10}^{\prime }$ reduces the
light Higgs mass in a manner similar to the
action of the bottom Yukawa contribution at large tan$\beta$~\cite{Brignole}.
Also, we require that the $(10,\overline{10})$ fields are $R$-parity odd.
We assume furthermore that mixing of vector-like particles with the SM
are small so as to not violate bounds on flavor changing processes.
Nevertheless, even a small mixing of this variety allows vector-like particle
from $(10,\overline{10})$ to have prompt decay and avoid cosmological problems. In the superpotential, $M_V$ defines the masses for  the vector-like fields, $Q_{10}$, $U_{10}$ and $E_{10}$. We will use $M_U$ and $M_D$ to represent $\kappa_{10}<v_{u}>$ and  $\kappa'_{10}<v_{d}>$ respectively.

\begin{figure}[t]
\centering \includegraphics[angle=0, width=12cm]{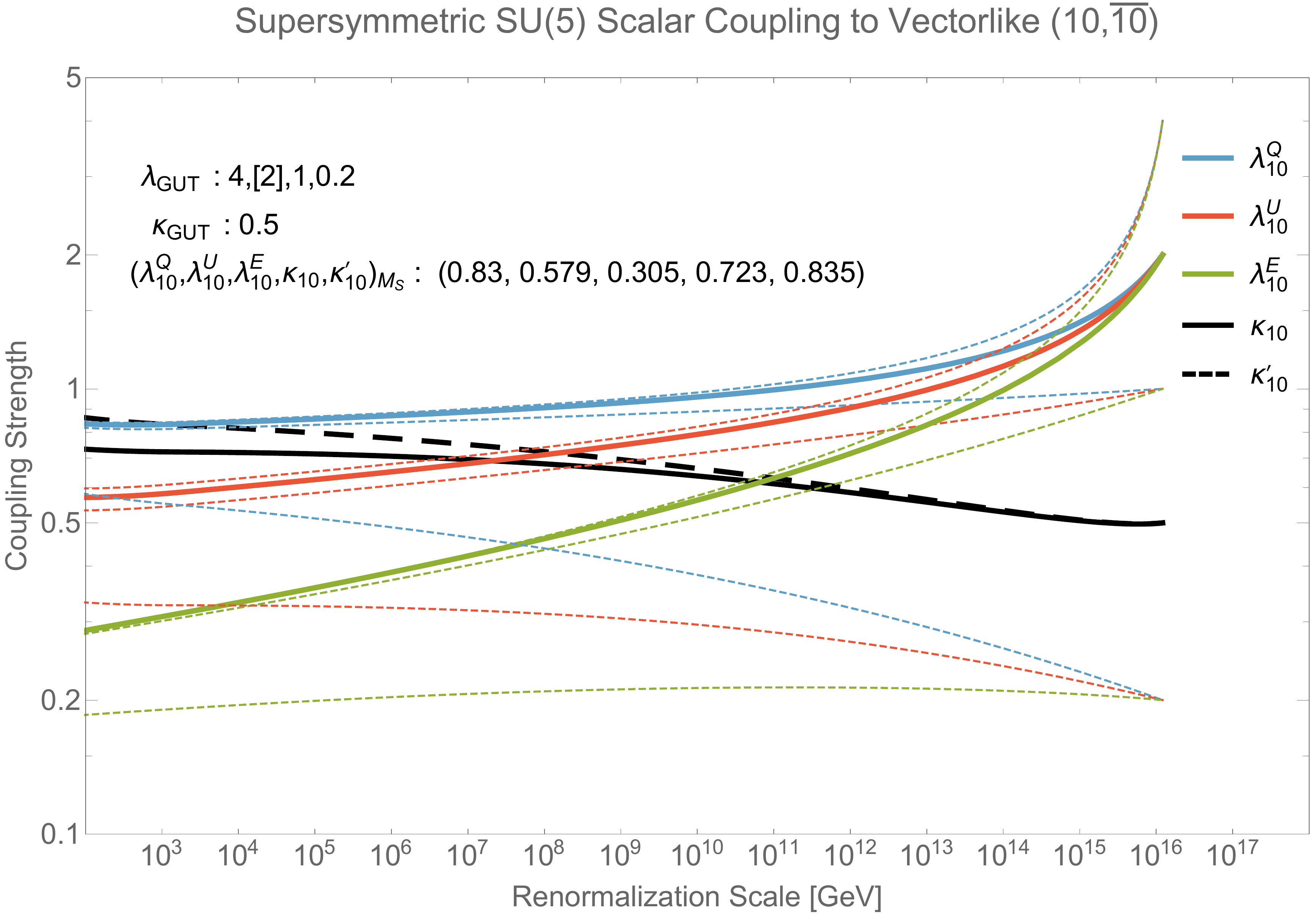}
\vspace{-0.3cm} \caption{Renormalization group evolution of the $(10,\overline{10})$
couplings presented in Eq.~(\ref{nn1}).
The $\kappa$ are analogs of the MSSM Yukawa couplings, linking the vector-like fields to the Higgs.
The $\lambda$ are couplings of vector-like fields and their conjugates to the scalar $S$.
The bold, solid curves, and also the corresponding numerical coupling values provided at $M_S$,
adopt the value $\lambda_{\rm GUT} = 2$.  Additional thin, dashed sets
of curves for the evolution of the $\lambda$ couplings from different boundary values
(4, 2, 1, and 0.2) are included in order to exhibit the infrared quasi-fixed point behavior.
Mass thresholds consistent with the application of $(M_V^{\rm GUT}, M_{\rm SUSY}) = (350~{\rm GeV}, 2~{\rm TeV})$ are employed.}
\label{yukawa10}
\end{figure}

\begin{figure}[t]
\centering \includegraphics[angle=0, width=12cm]{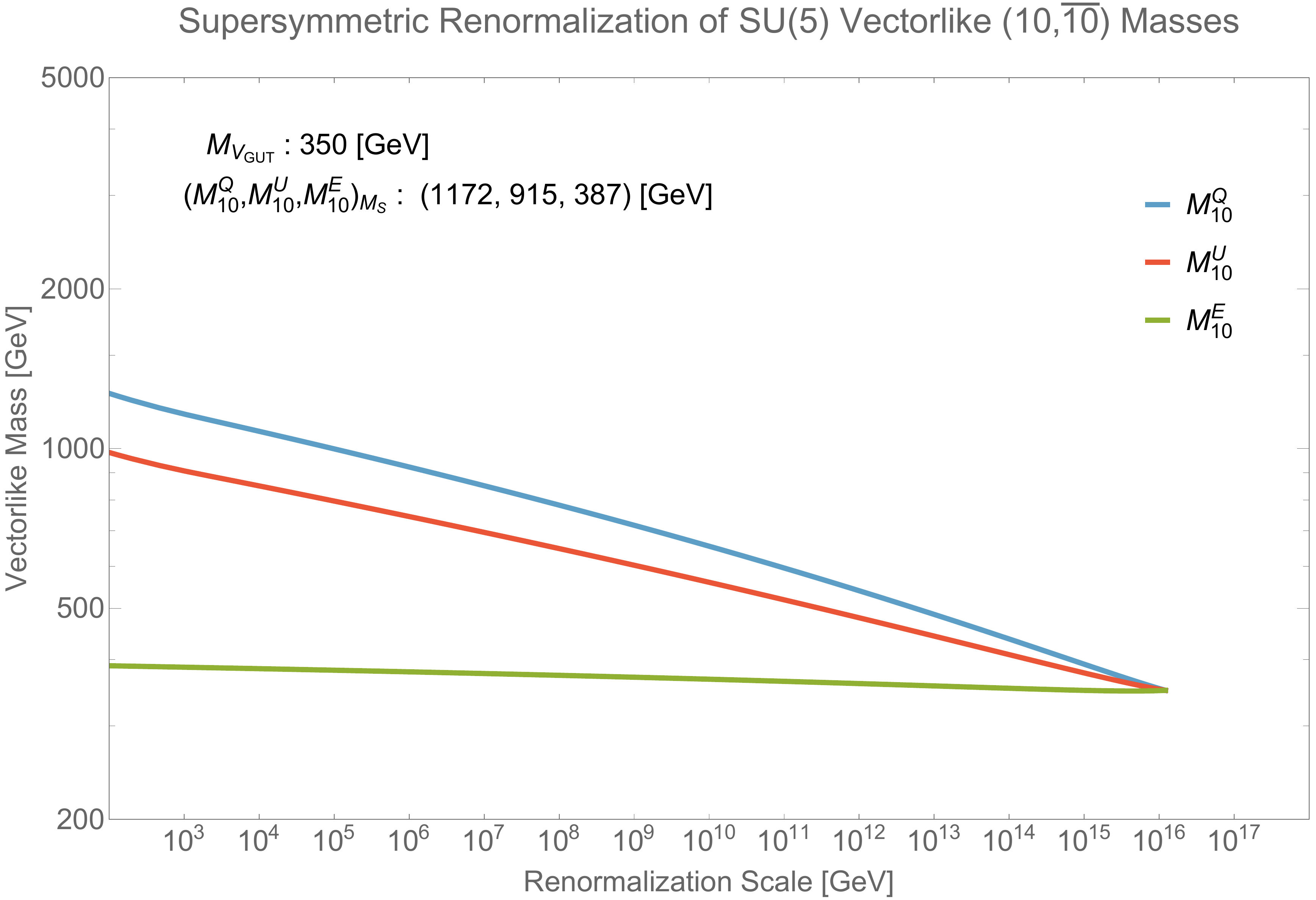}
\vspace{-0.3cm} \caption{Renormalization group evolution of the $(10,\overline{10})$
vector-like mass terms presented in Eq.~(\ref{nn1}).
The GUT-scale boundary values $(\lambda,\kappa,M_V^{\rm GUT}, M_{\rm SUSY}) = (2,0.5,350~{\rm GeV}, 2~{\rm TeV})$ are adopted.}
\label{mass10}
\end{figure}

In Figure~\ref{yukawa10}, we show the renormalization group evolution of the $(10,\overline{10})$
couplings in Eq.~(\ref{nn1}) from universal boundary values of $\lambda=4,2,1,0.2$ and $\kappa=0.5$ at $M_{\rm GUT}$.
A strong quasi-fixed point attraction in the infrared is observed for the $\lambda$s.
This behavior is analogous to that observed by Pendleton and Ross~\cite{Pendleton:1980as}
for the top quark Yukawa coupling, wherein attraction toward an approximately
universal value is observed at low energy
when starting from a sufficiently large coupling at the GUT-scale boundary.
There is no universal behavior for weak coupling at the GUT-scale, which generically
also results in weaker coupling at low energy.
In Figure~\ref{mass10}, we show the corresponding evolution of the
$(10,\overline{10})$ vector-like masses in Eq.~(\ref{nn1})
for a boundary value of $M_{V_{\rm GUT}}=350$~GeV. Although not exhibited in Figure~\ref{mass10}, a quasi-fixed point behavior of the vector-like particle masses, similar to the couplings, is also observed.  

Employing the effective potential approach we calculate the
additional contribution from the vector-like particles to the CP-even
Higgs mass at one loop level. A similar calculation was carried out
in Ref.~\cite{Gogoladze:2009bd,Babu:2004xg}.
\begin{eqnarray}
\left[ M_{H}^{2}\right] _{10} &=&-M_{Z}^{2}\cos ^{2}2\beta \left(
\frac{3}{8\pi ^{2}}\kappa_{10} ^{2}t_{V}\right)
+\frac{3}{4\pi ^{2}}\kappa_{10} ^{4}v^{2}\sin ^{2}\beta \left[ t_{V}+\frac{1%
}{2}X_{\kappa_{10} }\right],  \label{eq1}
\end{eqnarray}%
where we have assumed $M_{V}\gg M_{D}$ and
\begin{equation}
X_{\kappa_{10} }=\frac{4\widetilde{A}_{\kappa_{10} }^{2}\left(
3M_{\text SUSY}^{2}+2M_{V}^{2}\right) -\widetilde{A}_{\kappa_{10}
}^{4}-8M_{\text SUSY}^{2}M_{V}^{2}-10M_{\text SUSY}^{4}}{6\left(
M_{\text SUSY}^{2}+M_{V}^{2}\right) ^{2}} \label{f1}
\end{equation}%
\begin{equation}
t_{V}=\log \left( \frac{%
M_{\text SUSY}^{2}+M_{V}^{2}}{M_{V}^{2}}\right)\,. \label{mm7}
\end{equation}
Here $\widetilde{A}_{\kappa_{10}}=A_{\kappa_{10}}-\mu \cot \beta $,
$A_{\kappa_{10}}$ is the $Q_{10}-U_{10}$ soft mixing
parameter and $\mu$ is the MSSM Higgs bilinear mixing term.
$M_{\text SUSY}  \simeq \sqrt{m_{\tilde{Q}_{3}}\,m_{\tilde{U}_{3}^c}}$, where
$m_{\tilde{Q}_{3}}$ and $m_{\tilde{U}_{3}^c}$ are the stop left and
stop right soft SUSY breaking masses at low scale.

Next, we present the leading 1- and 2- loop
contributions to the CP-even Higgs boson mass in the MSSM
\cite{at, Carena:1995wu}%
\begin{eqnarray}
\left[ M_{H}^{2}\right] _{\text MSSM} &=&M_{Z}^{2}\cos ^{2}2\beta \left( 1-\frac{3%
}{8\pi ^{2}}\frac{m_{t}^{2}}{v^{2}}t\right)  \nonumber \\
&+&\frac{3}{4\pi ^{2}}\frac{m_{t}^{4}}{v^{2}}\left[ t+\frac{1}{2}X_{t}+\frac{%
1}{\left( 4\pi \right) ^{2}}\left(
\frac{3}{2}\frac{m_{t}^{2}}{v^{2}}-32\pi \alpha _{s}\right) \left(
X_{t}t+t^{2}\right) \right], \label{eq2}
\end{eqnarray}
where
\\
\begin{eqnarray}
t =\log \left(
\frac{M_{\text SUSY}^{2}}{M_{t}^{2}}\right),~
X_{t} &=&\frac{2\widetilde{A}_{t}^{2}}{M_{\text SUSY}^{2}}\left( 1-\frac{\widetilde{A}%
_{t}^{2}}{12M_{\text SUSY}^{2}}\right), \label{A1}
\end{eqnarray}%
with $\widetilde{A}_{t}=A_{t}-\mu \cot \beta $, where
$A_{t}$ denotes the left stop and right stop soft
mixing parameter.

The light Higgs mass is then expressed as
\begin{eqnarray}
M_H^2= \left[ M_{H}^{2}\right] _{\text MSSM} + \left[ M_{H}^{2}\right]
_{10}. \label{max_10}
\end{eqnarray}%
From Eq. (\ref{eq1}), we observe that the Higgs mass is very
sensitive to the value of $\kappa_{10}$, which cannot however be
taken arbitrary large without losing perturbativity of the theory up to $M_{\rm GUT}$.
We must therefore solve the RGE for $\kappa_{10}$
to make sure that it remains perturbative up to the GUT scale.
It was shown in Ref.~\cite{Gogoladze:2009bd} that
$\kappa_{10}\approx 1$ can successfully realize a 125 GeV Higgs mass without
invoking multi-TeV stop quark masses or a maximal value for the $A_t$ term.

Note that there is an additional tree-level contribution to the CP-even Higgs boson mass from the
$\lambda S\, H_u \, H_d$ coupling, given approximately by $\Delta M_H^2 \approx \lambda^2 v^2 \sin^2 2\beta$.
Because this contribution is significant only when $\tan\beta \approx 2$,
and $\lambda$ is around 0.5 -- 0.7, we will not consider this contribution further here.

\begin{figure}[t]
\centering \includegraphics[angle=0, width=10cm]{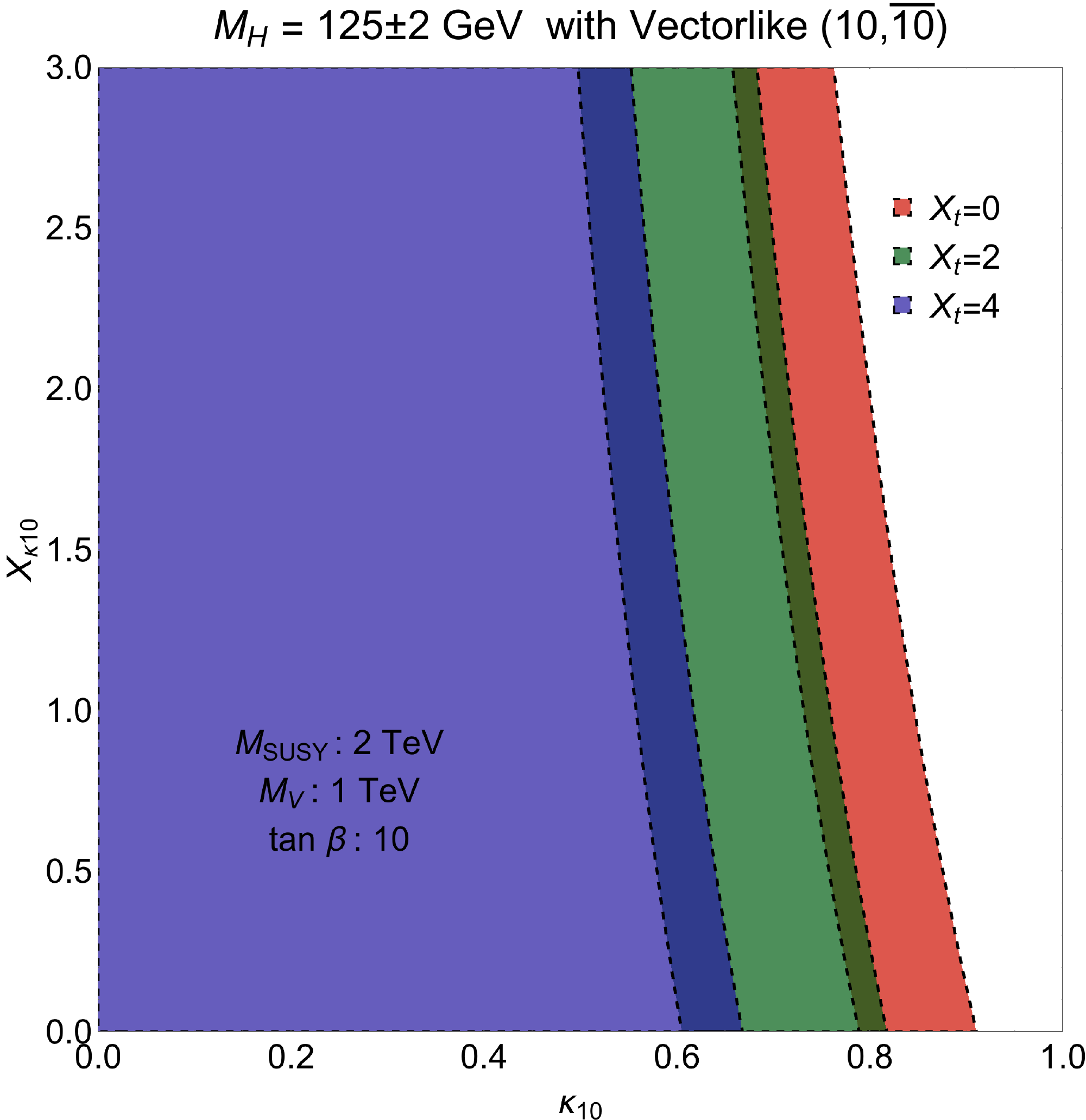}
\vspace{-0.3cm} \caption{
Regions of the parameter space for $\kappa_{10}$ of Eq.~(\ref{nn1})
and $X_{\kappa_{10}}$ of Eq.~(\ref{f1}) that are consistent
with $M_H=125\pm2$~GeV for various values of $X_t$ from Eq.~(\ref{A1}).
The darkened regions represent overlap between adjacent bands.}
\label{higgs10}
\end{figure}

In Figure~\ref{higgs10} we outline the viable parameter regions for $\kappa_{10}$ and $X_{\kappa_{10}}$,
given a Higgs mass of $125\pm2$~GeV, for $M_{\rm SUSY} = 2$~TeV and various values of $X_t$.
We see that the vector particle contribution can be
significant, allowing us to find the correct Higgs mass for smaller $X_t$.
The dependency on the $X_{\kappa_{10}}$ term, which depends strongly on the scale of the A-terms, is relatively weak.
The much stronger dependency is on the coupling $\kappa_{10}$.
Typically, for $\kappa_{10} < 1/2$, there is very little boost to the MSSM Higgs mass, although
the effect becomes substantial very quickly as this coupling goes to $3/4$ or higher.
Smaller couplings are more plausible if the SUSY scale is heavier and/or the
vector-like matter scale is lower.  Finally, the dependence on $\tan \beta$ is weak.

\subsection{{MSSM }+ $\mathbf{(5,\overline{5})}$}

The representation $(5,\overline{5})$ of $SU(5)$
decomposes under the MSSM gauge symmetry as follows:%
\begin{eqnarray}
  5 &=&{\overline L}_{5}\left( 1,2,\frac{1}{2}\right)
  + {\overline D}_{5}\left(3,1,-\frac{1}{3}\right)~,~\nonumber \\
  \overline{5} &=& L_{5}\left(1,2,-\frac{1}{2}\right)
   +D_{5}\left(\overline{3},1,\frac{1}{3}\right)~.
\end{eqnarray}%

By itself,  having only ($5,\overline 5$) does not allow for
any new Yukawa coupling to the MSSM Higgs unless the new states in the $\overline{5}$
are mixed with the usual $d^c$-quarks and lepton doublets. Such a
possibility is very
strongly constrained (by flavor violation and unitarity of the CKM
matrix, among others), and so we will suppress all such mixings.
However, if we introduce an
SM gauge singlet $S$, then Yukawa couplings of the form (in
$SU(5)$ notation) $\overline 5 \cdot S \cdot H_u$ and $5 \cdot S \cdot
H_d $ are permitted. Here we also introduce a singlet $S$-field, as in the previous section.
In this case the MSSM superpotential has the following additional contribution

\begin{eqnarray}
W \subset \kappa _{5}L_{5}{S}H_{u}+\kappa _{5}^{\prime }\overline{L}
_{5}S H_{d}+ \lambda^D_{5} S\, \overline{D}_{5} \,D_{5} + \lambda^L_{5} S \,\overline{L}
_{5}\, L_{5} + \lambda S \,H_{u}\, H_{d} + m_S S^2 + \nonumber  \\
M_{V}\left( S \overline S +
\overline{L}_{5}L_{5}+\overline{D_{5}}%
D_{5}\right).\label{dd66}
\end{eqnarray}
%
We also assume that there is an
additional symmetry forbidding mixing between the vector-like
particles and the MSSM matter fields. With this assumption the
singlet field $S$ cannot be identified with the right-handed sneutrino.

\begin{figure}[t]
\centering
\includegraphics[angle=0, width=12cm]{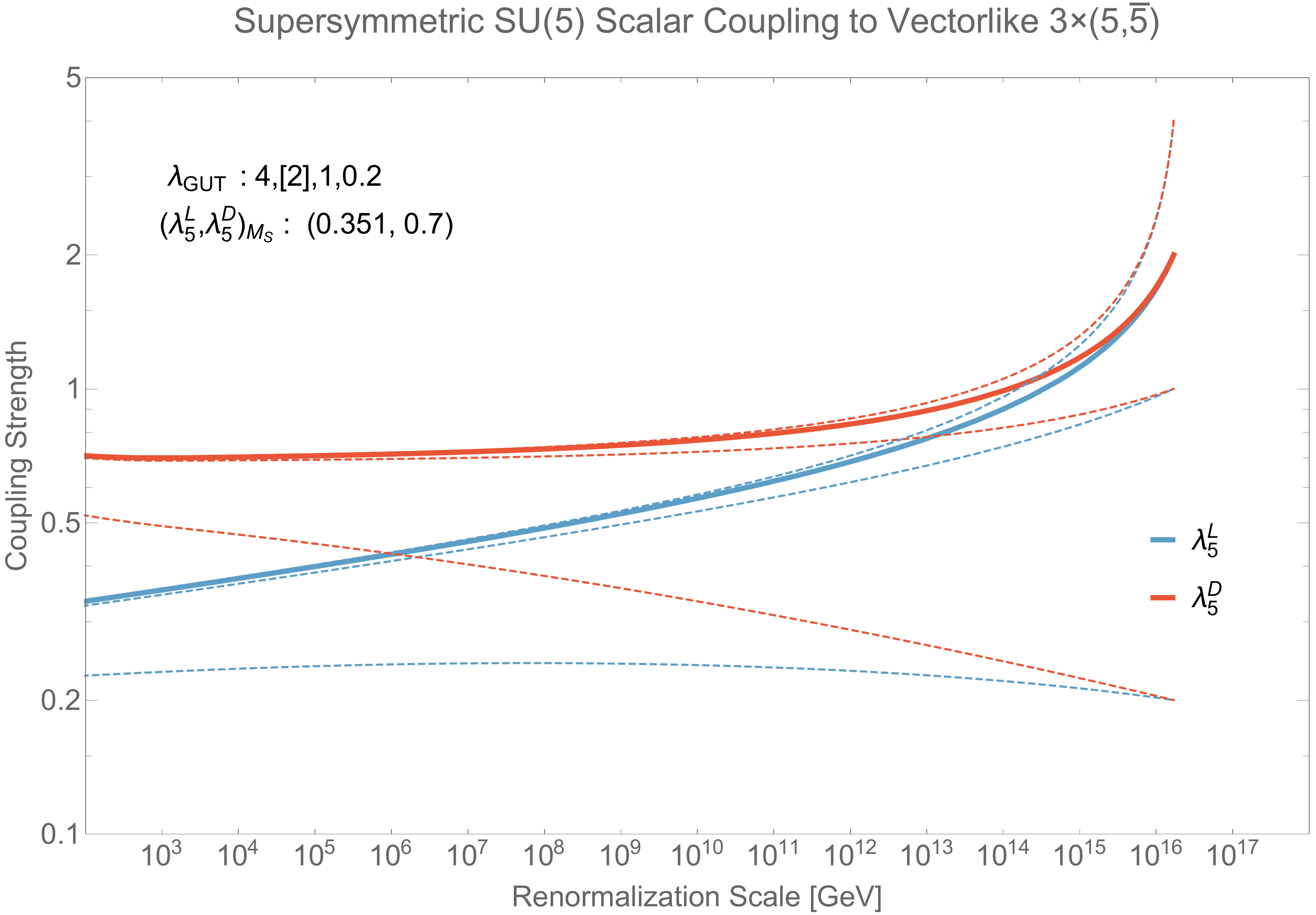}
\vspace{-0.3cm} \caption{Renormalization group evolution of the $(5,\overline{5})$
$\lambda$ couplings presented in Eq.~(\ref{dd66}), which
link the vector-like fields and their conjugates to the scalar $S$.
The bold, solid curves, and also the corresponding numerical coupling values provided at $M_S$,
adopt the value $\lambda_{\rm GUT} = 2$.  Additional thin, dashed sets
of curves for the evolution of the $\lambda$ couplings from different boundary values
(4, 2, 1, and 0.2) are included in order to exhibit the infrared quasi-fixed point behavior.
Mass thresholds consistent with the application of $(M_V^{\rm GUT}, M_{\rm SUSY}) = (350~{\rm GeV},2~{\rm TeV})$ are employed.}
\label{yukawa5}
\end{figure}

\begin{figure}[t]
\centering \includegraphics[angle=0, width=12cm]{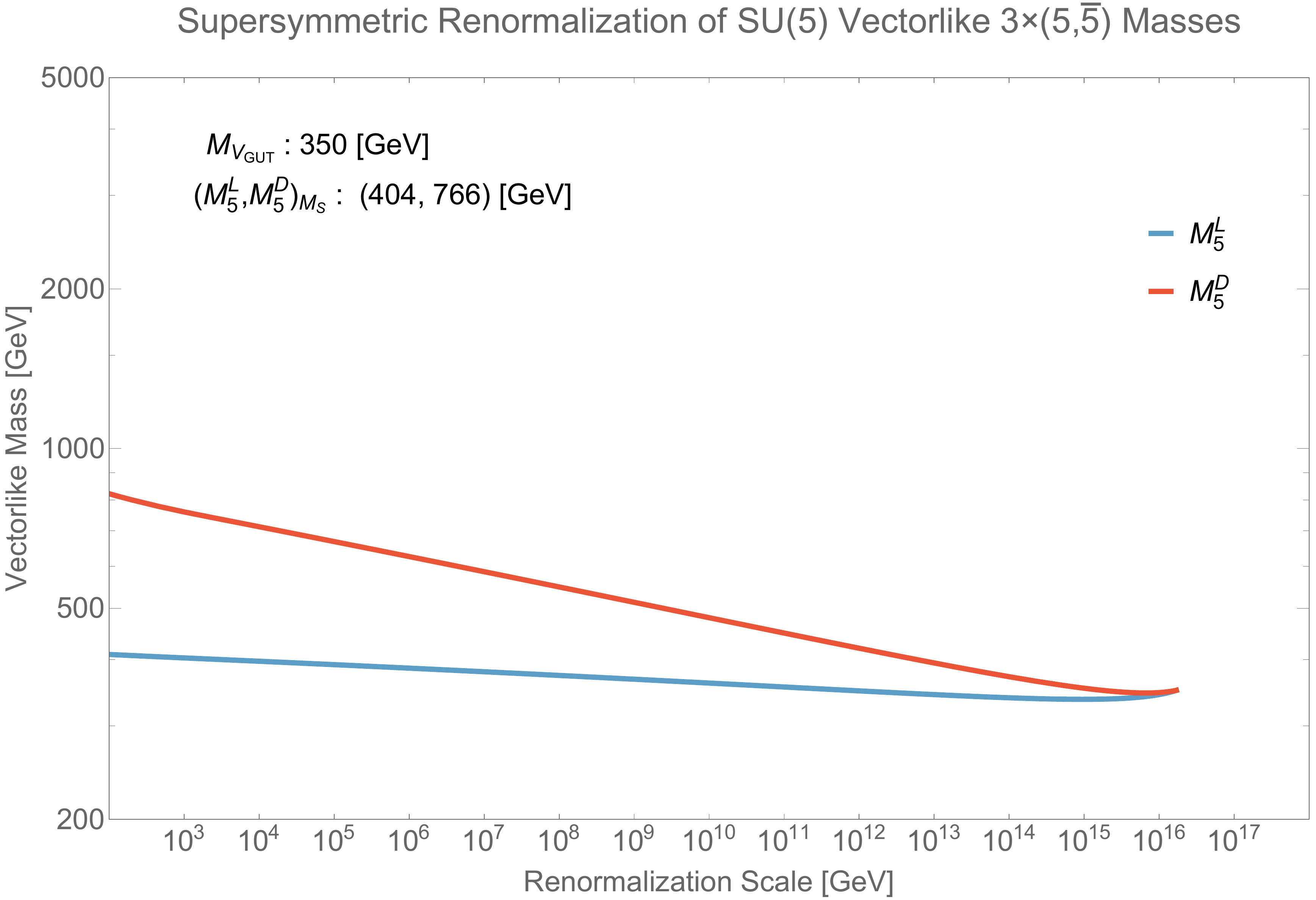}
\vspace{-0.3cm} \caption{Renormalization group evolution of the $(5,\overline{5})$
vector-like mass terms presented in Eq.~(\ref{dd66}).
The GUT-scale boundary values $(\lambda,M_V^{\rm GUT}, M_{\rm SUSY}) = (2,350~{\rm GeV},2~{\rm TeV})$ are adopted.}
\label{mass5}
\end{figure}

In Figure~\ref{yukawa5}, we show the renormalization group evolution of the $(5,\overline{5})$
couplings in Eq.~(\ref{dd66}) from universal boundary values of $\lambda=4,2,1,0.2$ at $M_{\rm GUT}$.
An infrared quasi-fixed point attraction for the $\lambda$s is again observed. 
In Figure~\ref{mass5}, we show the corresponding evolution of the
$(5,\overline{5})$ vector-like masses in Eq.~(\ref{dd66})
for a boundary value of $M_{V_{\rm GUT}}=350$~GeV. A quasi-fixed point behavior is true for masses in $(5,\overline{5})$ unification scheme as well.

Using the effective potential approach we calculate the additional
contribution to the CP-even Higgs mass at one loop~\cite{Gogoladze:2009bd}
\begin{eqnarray}
\left[ M_{H}^{2}\right] _{5} &=&-M_{Z}^{2}\cos ^{2}2\beta \left(
\frac{1}{8\pi ^{2}}\kappa_{5} ^{2}t_{V}\right)
+\frac{1}{4\pi ^{2}}\kappa_{5} ^{4}v^{2}\sin ^{2}\beta \left[ t_{V}+\frac{1%
}{2}X_{\kappa_{5} }\right],  \label{e3}
\end{eqnarray}%
where we have assumed $M_{V}\gg M_{D}$ and
\begin{equation}
X_{\kappa_{5} }=\frac{4\widetilde{A}_{\kappa_{5} }^{2}\left(
3M_{\text SUSY}^{2}+2M_{V}^{2}\right) -\widetilde{A}_{\kappa_{5}
}^{4}-8M_{\text SUSY}^{2}M_{V}^{2}-10M_{\text SUSY}^{4}}{6\left(
M_{\text SUSY}^{2}+M_{V}^{2}\right) ^{2}} \label{X3}
\end{equation}%
\begin{equation}
t_{V}=\log \left( \frac{%
M_{\text SUSY}^{2}+M_{V}^{2}}{M_{V}^{2}}\right),
\end{equation}
with $\widetilde{A}_{\kappa_{5}}=A_{\kappa_{5}}-\mu \cot \beta $,
where $A_{\kappa_{5}}$ is the $L_{5}-S$ soft mixing parameter and
$\mu$ is the MSSM Higgs bilinear mixing term.

\begin{figure}[t]
\centering \includegraphics[angle=0, width=10cm]{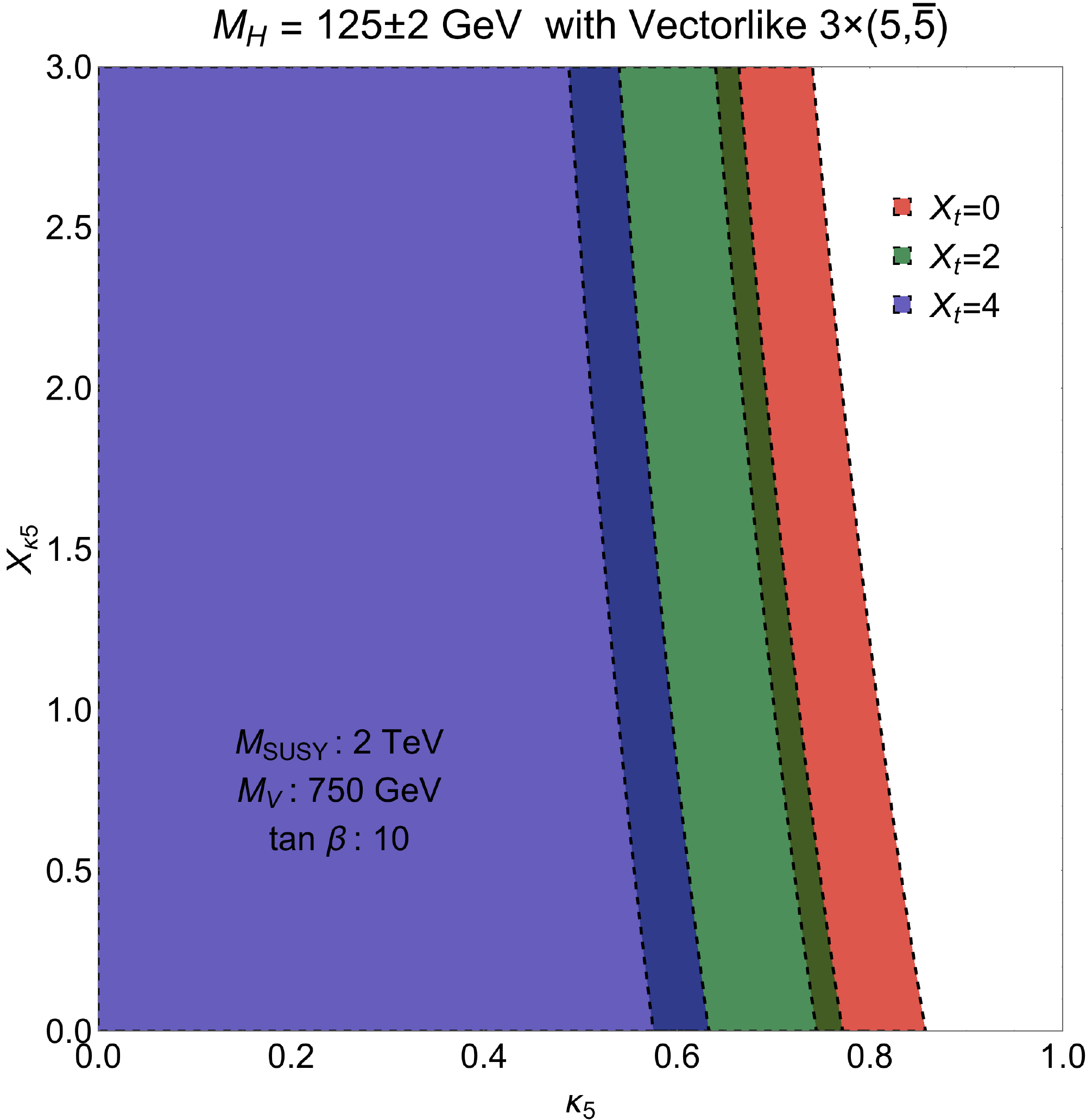}
\vspace{-0.3cm} \caption{
Regions of the parameter space for $\kappa_{5}$ of Eq.~(\ref{dd66})
and $X_{\kappa_{5}}$ of Eq.~(\ref{X3}) that are consistent
with $M_H=125\pm2$~GeV for various values of $X_t$ from Eq.~(\ref{A1}).
The darkened regions represent overlap between adjacent bands.}
 \label{higgs5}
\end{figure}

In Figure~\ref{higgs5} we outline the viable parameter regions for $\kappa_{5}$ and $X_{\kappa_{5}}$,
given a Higgs mass of $125\pm2$~GeV, for $M_{\rm SUSY} = 2$~TeV and various values of $X_t$.
We see that the vector particle contribution can be
significant and allows us to find the correct Higgs mass for smaller $X_t$.
The previous discussion of Figure~\ref{higgs10} carries over.

We also have a similar situation for the MSSM + $(16,\overline{16})$ case,
although perturbative gauge coupling unification suggests that the SUSY scale
should be pushed upward to several TeV.

\section{Neutral vector-like doublet component as a resonance }
\label{doublet}
In this section we present a new mechanism for the generation of a di-photon excess via resonance of a
neutral component of $L_5$.
Following standard notation, we denote the neutral and charged components of $L_5$ as $\nu_5$ and $E_5$.
Here we assume that we have a $(10,\overline{10} + 5,\overline{5})$ set of vector-like particles at low scale.
As mentioned above, in order to maintain successful gauge coupling unification in the MSSM
vector-like particles should compose full representations of $SU(5)$.
On the other hand, gauge coupling unification does not
require that all must come from the same representation of $SU(5)$.
In particular,  in the orbifold GUTs~\cite{kawa, GAFF, LHYN}
and F-theory GUTs~\cite{Donagi:2008ca, Beasley:2008dc, Beasley:2008kw, Donagi:2008kj}
(See Ref.~\cite{Li:2010hi} and references therein.), etc., we can split the multiplets,
and then the multiplets in the  $(10,\overline{10} + 5,\overline{5})$
can indeed arise from different $SU(5)$ representations.

\begin{figure}[t]
\centering
\includegraphics[angle=0, width=8cm]{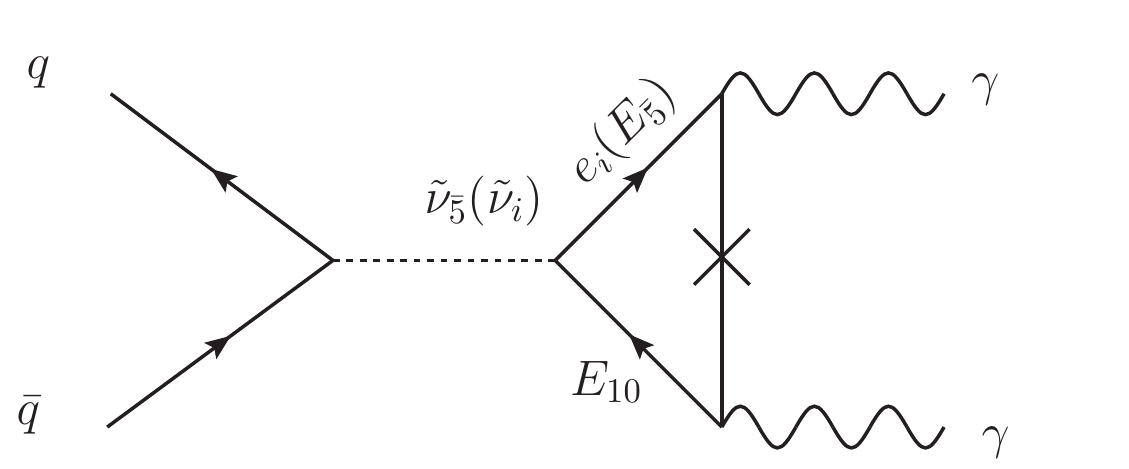}
\vspace{-0.3cm} \caption{Production and decay of the neutral component of $L_5$ via the $Q_{10} L_{5} D_{5}$ and
 $L_5 L_5 E_{10}$ couplings.} \label{figg12}
\end{figure}


In order to explain the diphoton excess we assume the existence of a $Z_3$ baryon parity~\cite{Ibanez:1991pr} in the theory.
So, in principle, the MSSM matter fields will mix with vector like fields.
In this framework the relevant couplings to the diphoton excess are the following:
\begin{eqnarray}
\label{baryon_parity}
  W \subset  \eta_{ijk} Q_i\, L_{j} \, D^c_k + \eta^D_{ij5} Q_i\, \overline{L}_{ 5} \, D^c_j + \eta_1 \, Q_{10}\, \overline{L}_{ 5} \, \overline{D}_{{5}} + \eta^{\prime}_{i} Q_{10}\, L_i \, \overline{D}_{ 5}\nonumber \\
  + \eta_i \overline{L}_{ 5}\, L_i\, E_{10}
 + \eta_{2} \overline{L}_{ 5}\, H_d\, E_{10}
  ,~
\end{eqnarray}%
where the fields with Latin indices are the SM fields. These couplings will lead to the Feynman diagram presented in Figure~\ref{figg12}.
The first four term in the above equation are relevant for the production, while the fifth term is relevant for the decay of $\tilde{\nu}_5$.
The last term is for mass insertion. Here the MSSM sneutrinos can mix with $\tilde{\nu}_5$ and with appropriate choice of parameters we can
have two neutral scalars with very close-by masses which may lead to two nearby resonances. It is interesting to note that in this model we
have lepton number violation, which can generate proper mass and mixing of the neutrinos~\cite{Chun:1998gp}.


Another way to generate the diagram presented in Figure~\ref{figg12} is to assume that all colored particles
$(Q_{10},\overline{Q}_{10}+U_{10},\overline{U}_{10}+D_{5},\overline{D}_{5})$
from $(10,\overline{10} + 5,\overline{5})$ are $R$-parity odd and
all colorless particles ($E_{10},\overline{E}_{10}+L_{5},\overline{L}_{5}$) are $R$-parity even.
In this case  we need at least two pairs of $(5,\overline{5})$.
Then, in addition to the interactions given in Eq.~\ref{dd66}, we have
\begin{eqnarray}
\label{two_doublets}
  W &\subset& \eta^U_{ijk} Q_i\, L^k_5 \, U_j+ \eta^D_{ijk} Q_i\, \overline{L}^k_5 \, D_j
  + \eta_{kl} \overline{L}^{k}_5\, \overline{L}^l_5\, E_{10} +  \overline{\eta}_{kl} {L}^{k}_5\, {L}^l_5\, \overline{E}_{10}
  \nonumber \\
 && + \eta_{k} \overline{L}^{k}_5\, H_d\, E_{10} +  \overline{\eta}_{k} {L}^{k}_5\, H_u\, \overline{E}_{10}
  ~,~
\end{eqnarray}%
  where $\eta_{kl}$ and $\overline{\eta}_{kl}$ are anti-symmetric in $k$ and $l$.
These couplings can explain
the di-photon excess observed at the LHC as follows: the first two operators facilitate production,
the third and fourth operators yield decays, and the last two operators provide for mass insertions.
Unlike the previous proposals~\cite{Ding:2015rxx, Allanach:2015ixl},
R-parity is preserved here.
Moreover, again with proper choice of parameters, we can make $L^k_5$ and $\overline{L}^k_5$ nearly degenerate and can explain
a large decay width around 45 GeV. In this case gauge coupling becomes non-perturbative before $10^{16}$ GeV.
We should point out that this scenario can be embedded nicely into orbifold GUT framework, which we will discuss in a follow-up paper.




\section{The Di-Photon Excess }

\begin{table}[h]
\begin{center}
\setlength\tabcolsep{6pt}
\begin{tabular}{|c|c|c|c|c|c|c|c|}
\hline \hline
 & $\kappa^{'}_1$ & $\kappa^{'}_2$ & $\kappa^{'}_3$ & & $\kappa^{'}_1$ & $\kappa^{'}_2$ & $\kappa^{'}_3$ \\
 \hline
~($Q$, ${\overline Q}$)~ & ~$\frac{\lambda^{Q}_{10} g_Y^2}{96\pi^2 M_Q}$~ &  ~$\frac{3 \lambda^{Q}_{10} g_2^2}{32\pi^2 M_Q}$~ &
 ~$\frac{\lambda^{Q}_{10} g_3^2}{16\pi^2 M_Q}$~ &
~($L$, ${\overline L}$)~ & ~$\frac{\lambda^{L}_{5} g_Y^2}{32\pi^2 M_L}$~ &  ~$\frac{ \lambda^{L}_{5} g_2^2}{32\pi^2 M_L}$~ &
 ~0~\\
\hline
~($U$, ${\overline U}$)~ & ~$\frac{\lambda^{U}_{10} g_Y^2}{12\pi^2 M_U}$~ &  0 &
 ~$\frac{\lambda^{U}_{10} g_3^2}{32\pi^2 M_U}$~ &
~($E$, ${\overline E}$)~ & ~$\frac{\lambda^{E}_5 g_Y^2}{16\pi^2 M_E}$~ &  0 &
~0~\\
\hline
~($D$, ${\overline D}$)~ & ~$\frac{\lambda^{D}_5 g_Y^2}{48\pi^2 M_D}$~ &  0 &
 ~$\frac{\lambda^{D}_5 g_3^2}{32\pi^2 M_D}$~ &
 &  &   &
 ~~\\
\hline \hline
\end{tabular}
\caption{The coefficients $\kappa^{'}_i$ ($i=1,2,3$) for different vector-like particles.
Note that the effective couplings $\kappa_i$ can obtained from the above coefficients by
multiplying them by the loop functions $A_{1/2}(\tau_F)$ (for $F=Q_{10},U_{10},D_{5},L_{5},E_{10}$) and
$A_{0}(\tau_{\tilde{F}})$ (for SUSY partners of $F$), presented in Eq.~\ref{loop_function_1}}
\label{T1Ki}
\end{center}
\end{table}

The heavy $F=Q_{10},U_{10},D_{5},L_{5},E_{10}$ fermions, as well as their supersymmetric scalar partners, can induce effective
loop-level couplings between $S$ and the SM gauge bosons, as given in Table~\ref{T1Ki}.
Likewise, couplings to the fermion(s) $N$ can lead to invisible tree-level decays at the collider whenever kinematically allowed,
\bea
\label{eq:Leff}
\mathcal{L}_{eff.} =  \kappa_1 S B_{\mu \nu} B^{\mu \nu} + \kappa_2 S W^{j}_{\mu \nu} W^{j \mu \nu} +  \kappa_3 S G^{a}_{\mu \nu} G^{a \mu \nu}  +\kappa_{NN}S \overline{N}N,
\eea
where $B_{\mu \nu}$, $W^{j}_{\mu \nu}$ and $G^{a}_{\mu \nu}$ respectively represent the field strength tensors of
the SM $U(1)_Y$, $SU(2)_L$ and $SU(3)_c$ groups, while $j=1,2,3$ and $a=1,2\ldots8$ are indices
of the associated non-Abelian adjoint representations.  The effective couplings $\kappa_i$ ($i=1,2,3$)
can be obtained from the coefficients $\kappa^{'}_i$ presented in Table~\ref{T1Ki} by,

\bea
\kappa_i =   \sum_{n=1}^k \kappa^{'}_{i_n} \, A_{1/2}(\tau_{F_n})  + \sum_{n=1}^{2k} \kappa^{'}_{i_n} \, \dfrac{M_{F_n} A_{F_n}}{2 M_{\tilde{F}_n}^{2}} A_{0}(\tau_{\tilde{F}_n})~,
\label{SUSY_interference}
\eea
where the summation extends over the number of copies of ($5, \, \overline{5}$) and ($10, \, \overline{10}$), and $A_F$ are the trilinear couplings of $S$ with the SUSY partners of the vector-like fermions.
The loop functions $A_{1/2}(\tau)$ and $A_{0}(\tau)$ are given by
\bea
A_{\frac{1}{2}}(\tau) & = & 2 \int_{0}^1 dx \int_{0}^{1-x} dz \, \frac{1-4xz}{1-xz\tau}, \nn \\
A_{0}(\tau) & = &  \int_{0}^1 dx \int_{0}^{1-x} dz \, \frac{4xz}{1-xz \tau},
\label{loop_function_1}
\eea
with $\tau = \frac{M_{S}^2}{M^2}$. Please note that in the decay width calculations involving massive gauge bosons, the effect of gauge boson mass on loop functions have been neglected since they change the loop functions only by $\sim 5 \%$. In addition we also assumed that the mixing between the sparticles ($\tilde{F}_n$) is negligible  in the formulas of Eq.~(\ref{SUSY_interference}).

After rotation to the physical gauge boson states, these effective couplings can be written for both isosinglet and $SU(2)_L$ doublet as,
\bea
 \kappa_{\gamma \gamma } & = & \kappa_1 \cos^{2} \theta_W + \kappa_2 \sin^{2} \theta_W \, , \nn \\
 \kappa_{Z Z } & = &  \kappa_2 \cos^{2} \theta_W + \kappa_1 \sin^{2} \theta_W \, , \nn \\
 \kappa_{Z \gamma } & = &  (\kappa_2 - \kappa_1) \sin 2 \theta_W \, , \nn \\
 \kappa_{WW } & = & 2 \kappa_2 \, , \nn\\
 \kappa_{gg } & = & \kappa_3 \, ,
\label{eq:eff_couplings_singlets}
\eea
where $\theta_W$ is the weak mixing angle.

The current LHC bounds on vector-like quark masses range from 735 GeV for $D$-type isosinglets
to 855 GeV for the doublet $Q$ (see~\cite{ATLAS-VLQ} and references therein). However, these bounds depends on BRs of the vector-like quarks to the relevant final states. The BRs for a particular BP depends on the mixing of vector-like quarks with SM  quarks and we can tune the mixing parameters to
satisfy the bounds.
The loop contribution from charged sfermions interfere constructively with those from fermions.
For simplicity, we assume a common mass for the heavy fermions and their superpartners, $M_F=M_{\tilde{F}}$,
during the evaluation of loop functions. With a reasonable choice of $A_{F}/M_{\tilde{F}}$,
including the sfermions enhances $\sigma_{\gamma\gamma}$ significantly. It is shown in Ref.~\cite{Babu:2004xg} that the requirement that there be
no charge- or color-breaking minima in the scalar potential
lower than the SM minimum in the presence of vector-like fields requires $A_F^2<6 (M_V^2+M_{\rm SUSY}^2)$. For the rest of the paper we select $A_{F}/M_{\tilde{F}}$ based on this upper limit of $A_F$ to maximize $\sigma_{\gamma \gamma}$.



At the LHC, the diphoton production cross-section by means of gluon gluon fusion can be written in the narrow-width approximation,
\be
  \sigma_{\gamma\gamma}  = K \dfrac{ \, \pi^2}{8 M_S}\dfrac{\Gamma(S\rightarrow gg)\Gamma(S \rightarrow \gamma\gamma)}{ \Gamma_S} \times 
 \dfrac{1}{s}\int \limits dx_1 dx_2  f_{g}(x_1) f_g(x_2) \delta \left(x_1 x_2 - \frac{M^{2}_S}{s} \right),
 \label{eq:diphoton_xsec}
\ee
where $f_g$ is the gluon parton distribution function inside a proton, $x$ denotes the fraction of each beam's energy
carried away by the corresponding gluon, $K$ is the QCD K-factor and $\sqrt{s}=13$ TeV.
$\Gamma_S = \Gamma_{\gamma \gamma}+\Gamma_{Z \gamma}+\Gamma_{Z Z}+\Gamma_{WW}+\Gamma_{gg}+\Gamma_{N \overline{N}}$
denotes the total decay width of $S$. We have used the PDFs of {\tt MSTW2008LO}~\cite{MSTW2008} for the
gluon luminosity calculation with the factorization scale set at $M_S$. We evaluated $\alpha_s$ to be 0.092 at our scale of interest but we found that $\alpha$ does not change significantly from its value (0.0078) at $M_Z$. A K-factor of 2.5 is used in our calculation, which is the K-factor for 750~GeV SM-like Higgs~\cite{K-factor}. We also included $\alpha_{s}^4$ correction to $\Gamma_{gg}$, which increases it by a factor of $\sim 1.7$~\cite{Chetyrkin:1997iv}.
Similarly, the diphoton cross-section due to $q\overline{q}$ annihilation in the narrow-width approximation is,
\be
  \sigma_{\gamma\gamma}  = K \dfrac{4 \, \pi^2}{9 M_S}\dfrac{\Gamma(S\rightarrow q \overline{q})\Gamma(S \rightarrow \gamma\gamma)}{ \Gamma_S} \times 
 \dfrac{1}{s}\int \limits dx_1 dx_2 \bigg\{ f_{q}(x_1) f_{\overline{q}}(x_2) \delta \left(x_1 x_2 - \frac{M^{2}_S}{s} \right) \, + \, 1 \leftrightarrow 2 \bigg\},
 \label{eq:diphoton_xsec_qq}
\ee
where $\Gamma_{gg}$ is replaced by $\Gamma_{q \overline{q}}$ in $\Gamma_S$ and  $f_q (f_{\overline{q}})$ represents 
the quark (anti-quark) parton distribution function inside a proton.
For the $q\overline{q}$ resonance we used a K-factor of 1.3, which is the K-factor for Drell-Yan production processes~\cite{K-factor:DY}.


We should emphasize here that the experimentally observed width of the resonance is quite large.
ATLAS reported a width as large as $\Gamma = 0.06 M_S$. However the data collected so far is insufficient to claim such a
broad width conclusively. Ref.~\cite{Falkowski} performed a likelihood analysis to fit both CMS and ATLAS data and
checked for their consistency against the 8 TeV data as well. Their fit to the combined run-I and run-II data indicates
that the resonance at 750 GeV can be fit by $\sigma_{\gamma \gamma} \sim 0.7 - 16$ fb for $\Gamma_S \sim 5-100$ GeV at $2\sigma$ level.
Nevertheless, we adopt a conservative point of view and take the indicated width at face value,
thus restricting our study to $\Gamma_S \sim 5-45$ GeV. {After the re-analysis of the data by CMS and ATLAS, Ref.~\cite{Nilles} has  performed a updated likelihood analysis to fit both 8 and 13 TeV datasets of both experiments. They conclude that a narrow (broad) width resonance between $\sim 730-755 \,\,  (710-765)$ GeV can be fit by $\sigma_{\gamma \gamma} \sim 1 - 5 \,\, (2.5 -9)$ fb at $2\sigma$ level, with the best-fit being at 2.6 (6.9) fb.}
We perform a new analysis on the relative likelihood of small or large widths in the case of dual
resonances centered on a mass $M$ with a gap $\Delta M$ in Section~\ref{sec:dual}.

\begin{figure}[t]
\centering \includegraphics[angle=0, width=7cm]{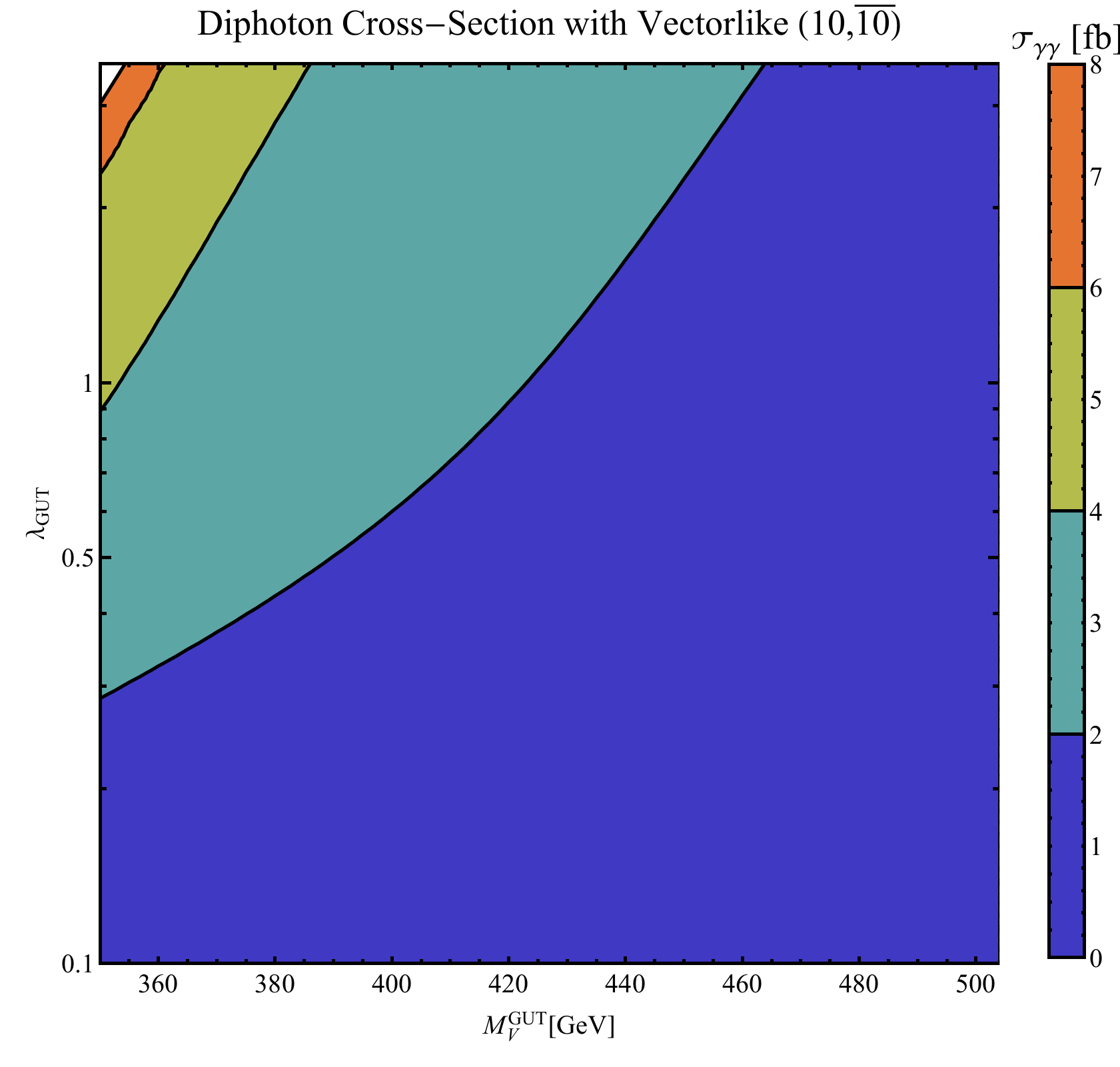}
\centering \includegraphics[angle=0, width=7cm]{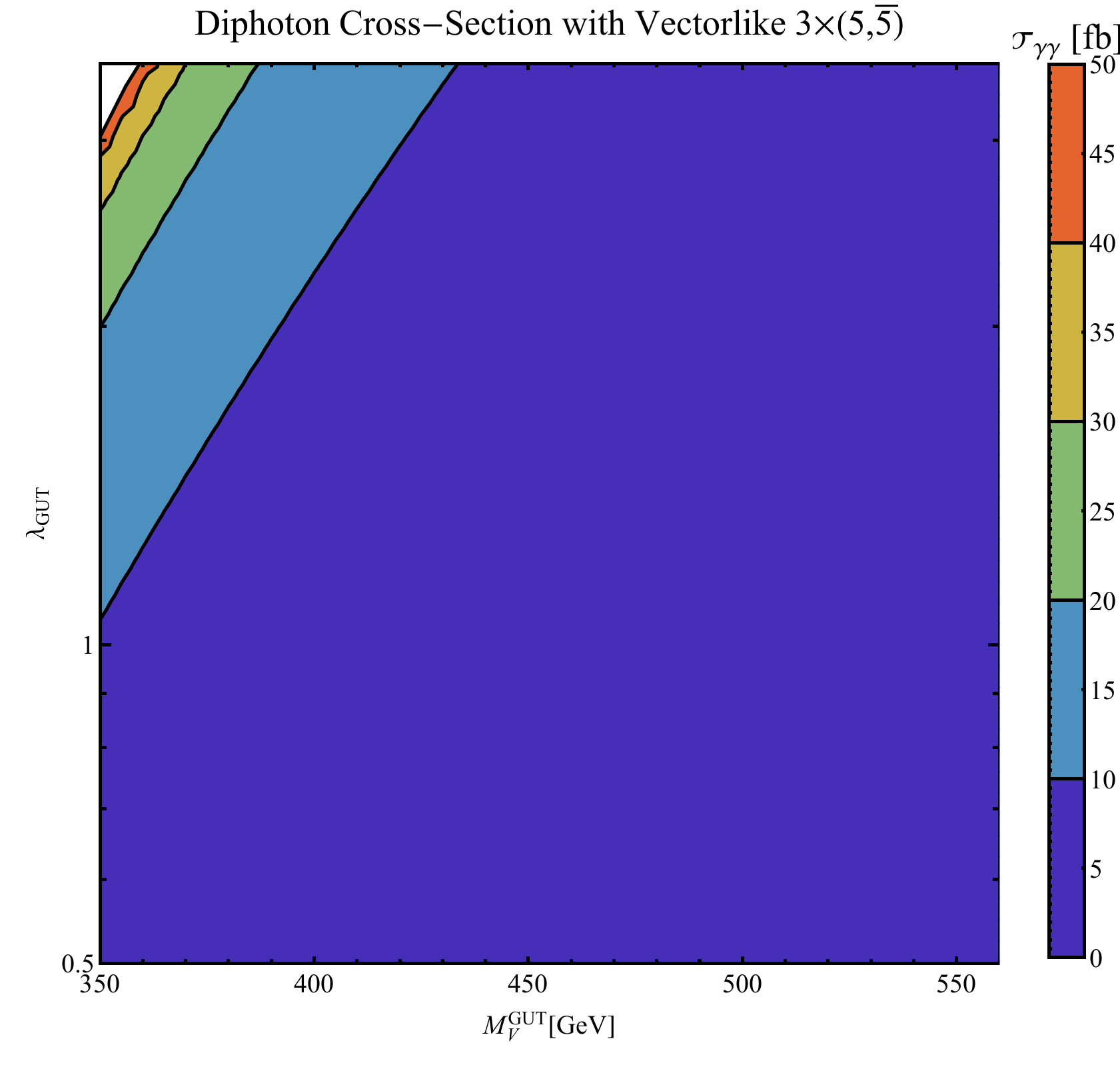}
\vspace{-0.3cm} 
\caption{Diphoton cross-section for different combinations of $M_{V}^{GUT}$ and $\lambda_{GUT}$ for both $(10,\overline{10})$ [left panel] and $3 \times (5,\overline{5})$ [right panel]. We fixed $M_{\rm SUSY}$ at 2 TeV and used the maximum value of $A_F$ allowed by color- and charge-breaking condition, $A_F^2<6 (M_V^2+M_{\rm SUSY}^2)$.}
\label{diphoton_xsec}
\end{figure}

The evolution of the couplings and masses of vector-like fermions
between the GUT scale and the scale of the observed resonance $M_S \sim 750$~GeV have been discussed in previous sections for both $(10,\overline{10})$ and $(5,\overline{5})$ extensions of the MSSM. We find that at least three copies of $(5,\overline{5})$ vector-like multiplets are needed in order to enhance the scalar
resonance cross-section to fit the data. Four copies of $(5,\overline{5})$ provide comparatively better fit to the
data, although this scenario, like the $(16,\overline{16})$, is on the edge of criticality and may exhibit a Landau pole if the vector-like and/or SUSY scales are too light. Similarly, two copies of $(10,\overline{10})$ multiplets fit the excess better than one copy, although this scenario is strictly incompatible with perturbative unification. In Fig.~\ref{diphoton_xsec}, we show diphoton cross-section for different combinations of $M_{V}^{GUT}$ and $\lambda_{GUT}$ for both $(10,\overline{10})$ and $3 \times (5,\overline{5})$. We fixed $M_{\rm SUSY}$ at 2 TeV and used the maximum value of $A_F$ allowed by color- and charge-breaking condition discussed earlier in the section. From Fig.~\ref{diphoton_xsec} we can see that for a range of values of $(M_{V}^{GUT},\lambda_{GUT})$, $\sigma_{\gamma \gamma} \lesssim 10$ fb and 50 fb can be achieved for $(10,\overline{10})$ and $3 \times (5,\overline{5})$ respectively. We did not consider any invisible decay width for the plots of Fig.~\ref{diphoton_xsec}. If we increase $M_{\rm SUSY}$ to 3 TeV the corresponding largest $\sigma_{\gamma \gamma}$ that can be attained are $\sim 12$ fb and 70 fb. One should infer that points for which $\sigma_{\gamma \gamma}$ are too high might be ruled out by the dijet constraint. However those points can survive if we introduce large invisible width and attain a smaller $\sigma_{\gamma \gamma}$ in the process.

It should be noted that the loop induced diphoton and dijet widths are inadequate to account for $\mathcal{O}(10)$ GeV width. 
The width associated with a resonance for both $3 \times (5,\overline{5})$ and $(10,\overline{10})$ cases are $\lesssim 1$ GeV.
Consequently, we require the width $\Gamma_{N \overline{N}}$ to be significant to obtain large width. In Table~\ref{tab:benchmarks},
we show the benchmark points for each unification scheme that explains the diphoton excess, and the
cross-sections into several leading associated final states.


\begin{table}[t]
\begin{center}
\setlength\tabcolsep{6pt}
\begin{tabular}{|c|c|c|c|}
\hline \hline
BP & BP-1 & BP-2 & BP-3  \\
\hline
& \multicolumn{1}{|c|}{$(10,\overline{10})$} & \multicolumn{2}{|c|}{$3 \times (5,\overline{5})$} \\
\hline
$M_V^{GUT}$ [GeV] & 425 &  455 & 360  \\
$\lambda_{GUT}$ [GeV] & 3.02 & 0.76 & 3.16 \\
\cline{2-3} 
$M_{\rm SUSY}$ [GeV] & \multicolumn{2}{|c|}{2000} & 3000  \\
\hline
$M_Q$ [GeV] & 1389 & -- & --   \\
$M_U$ [GeV] & 1086 & -- & -- \\
$M_D$ [GeV] & 456  & -- & --  \\
$M_E$ [GeV] & -- & 606 & 381 \\
$M_L$ [GeV] & --  & 1166 & 704 \\
\hline
$\lambda_Q$  & 0.83  & -- & --  \\
$\lambda_U$  & 0.60  & -- & --  \\
$\lambda_D$  & 0.31  & -- & --  \\
$\lambda_E$  & --  & 0.34 & 0.36 \\
$\lambda_L$  & --  & 0.68 & 0.69 \\
\hline
$A_{F}/M_{\tilde{F}}$ & 4.29 & 4.86 & 10.7 \\
$\Gamma_S$ [GeV] & 0.04 & 0.05 & 35.0 \\
\hline
$\sigma_{\gamma \gamma}$ & 2.60 & 2.60 & 1.00 \\
$\sigma_{ZZ}$ & 3.80 & 4.71 & 2.22 \\
$\sigma_{Z \gamma}$ & 0.24 & 0.56 & 0.40 \\
$\sigma_{WW}$ & 9.34  & 12.4 & 6.14 \\
$\sigma_{gg}$ & 505  & 576 & 71.4 \\
$\sigma_{invisible}$ & -- & -- & 5743 \\
\hline\hline 
\end{tabular}
\caption{Total decay width of $S$ and cross-section at $\sqrt{s} = 13$~TeV in associated final states. The invisible cross-section $\sigma_{N \overline{N}}$ is used to evaluate its monojet signal rate. The couplings and masses of vector-like fermions, as well as ($M_V^{GUT}, \lambda_{GUT}$), used are shown in Table}
\label{tab:benchmarks}
\end{center}
\end{table}

\begin{table}[h]
\begin{center}
\setlength\tabcolsep{6pt}
\begin{tabular}{|c|c|c|c|c|c|c|c|c|c|}
\hline\hline
 & & $\eta_{115}$ & $\Gamma_S$ & $\sigma_{\gamma \gamma}$ &  $\sigma_{Z Z}$ &  $\sigma_{Z \gamma}$ & $\sigma_{WW}$ & $\sigma_{q\overline{q}}$ & $\sigma_{invisible}$  \\
 & & & [GeV] & [fb] & [fb] & [fb] & [fb] & [fb] & [fb]  \\
 \hline
 \multirow{2}{*}{$\tilde{\nu}_{\overline{5}}$ resonance} & BP-4 & 0.07 & 0.22 & 1.20 & 0.10 & 0.69  & -- & 2332  & -- \\
                                      & BP-5 & 0.08 & 0.35 & 1.00 & 0.08 & 0.57 & -- & 2543 & 213 \\
 \hline
 \hline
\end{tabular}
\caption{Total decay width of $\tilde{\nu}_{\overline{5}}$ and cross-section at $\sqrt{s} = 13$~TeV in associated final states,
when $\tilde{\nu}_{\overline{5}}$ can act as the observed resonance (also includes $\tilde{\nu}^{*}_{\overline{5}}$
degenerate with $\tilde{\nu}_{\overline{5}}$), as discussed in Section~\ref{doublet}. We assumed that $\tilde{\nu}_{\overline{5}}$
couples to only the first generation quarks and the relevant coupling $\eta_{115}$ is also presented. $M^{E}_{10}$ and $\lambda^{E}_{10}$
are fixed at 400 GeV and 1 respectively for the evaluation of the loop functions. $\sigma_{invisible}$ stands for any invisible or
semi-invisible decay of $\tilde{\nu}_{\overline{5}}$ in SUSY particles. For dijet cross-section calculations the contribution from the
charged scalars of the doublet are also included.}
\label{tab:benchmarks2}
\end{center}
\end{table}

Table~\ref{tab:benchmarks} presents predicted cross-sections in leading final states for various restrictions on the total width $\Gamma_S$. 
BP-1 and BP-2 are two BPs that show the best-fit diphoton cross-section (2.6 fb) of Ref.~\cite{Nilles} for a narrow width resonance in the $(10,\overline{10})$ and $3\times(5,\overline{5})$ unification schemes. For $(10,\overline{10})$ we can not achieve a width larger than the diphoton resolution ($\sim 10 $ GeV) at 750 GeV. Hence we do not show any large width BP for $(10,\overline{10})$. However, as large a width as $\Gamma_S \sim 35$ GeV is possible for $3\times(5,\overline{5})$ scheme with a reasonable choice of $M_{\rm SUSY} \leq 3$ TeV. One such BP (BP-3) is shown in Table~\ref{tab:benchmarks}.

Next, let us discuss the constraints from few associated diboson ($S \rightarrow W^+ W^-, \,  ZZ, \, Z \gamma$) final states,
which arise from the couplings presented in Eq.~\ref{eq:Leff}.
The $W^+W^-, \, ZZ, \, Z\gamma$ signals are estimated to be at comparable rate to $\gamma\gamma$ channel as they originate
from the same set of couplings, as shown in Table~\ref{tab:benchmarks}. Among these three weak-boson channels, $Z\gamma$
is the most stringent and CMS~\cite{Khachatryan:2014rwa} constrains a monophoton signal to be less than $30$ fb with
missing energy \met $> 250$ GeV.  The supersymmetric $(10,\overline{10})$ and $3 \times (5,\overline{5})$
cases we considered here clearly satisfy these bounds.

Since the $gg$ and $\overline{N}N$ can take up sizeable partial width in comparison to $\gamma\gamma$, they should be
investigated more thoroughly. In their most recent dijet analysis, using 13 TeV data, both CMS~\cite{CMS-dijet13}
and ATLAS~\cite{ATLAS-dijet13} set a bound on dijet resonance mass only above 1 TeV.
However, CMS places a $2$ pb
bound on a 750 GeV $gg$ resonance from run-1 data~\cite{CMS:2015neg}. From this result we can easily estimate a
model independent bound on the relative ratio between $gg$ and $\gamma\gamma$,
\be
\frac{\text{BR}_{gg}}{\text{BR}_{\gamma\gamma}}  < \eta \cdot \frac{\sigma_{jj}^{\text{\scriptsize 8TeV}}}{\sigma_{\gamma\gamma}} \, ,
\ee
where $\eta=\sigma^{S}_{\text{13TeV}}/\sigma^{S}_\text{8TeV} \approx 5$ accounts for the difference in the $S$
production cross-section at 8 TeV. This constraint can rule out heavy
quark only models, where the two gluon channel dominates over diphoton due to the quarks' fractional electric charge.
In our unification models, the inclusion of heavy leptons in $(10,\overline{10})$ and $3 \times (5,\overline{5})$ enhances $\kappa_{\gamma\gamma}$
and becomes consistent with this dijet constraint. The dijet cross-sections of our BPs at 8 TeV, in comparison with the same at 13 TeV, are shown in Table~\ref{dijet8v13}.

\begin{table}[h]
\begin{center}
\setlength\tabcolsep{6pt}
\begin{tabular}{|c|c|c|}
\hline
\hline
BP & $\sigma^{\text{13 TeV}}_{gg/q\overline{q}}$ & $\sigma^{\text{8 TeV}}_{gg/q\overline{q}}$ \\
   & [fb] & [fb] \\
\hline
BP-1 & 505 & 107 \\
BP-2 & 576 & 122   \\
BP-3 & 71.4 & 15.1  \\
\hline
BP-4 & 2332 & 910 \\
BP-5 & 2543 & 992 \\
\hline
\hline
\end{tabular}
\caption{Comparison of dijet cross-section of our BPs at 8 TeV and 13 TeV.
The CMS 8 TeV bound on cross-section at 750 GeV dijet invariant mass is 2 (1) pb (for $gg$ ($q\overline{q}$) resonance)~\cite{CMS:2015neg}.}
\label{dijet8v13}
\end{center}
\end{table}

As it is also relevant to the dijet constraints, we should presently discuss the case where the neutral component of
a vector-like doublet ($\tilde{\nu}_{\overline{5}}$) acts as the resonance. We assume that $\tilde{\nu}^{*}_{\overline{5}}$ is
nearly degenerate with $\tilde{\nu}_{\overline{5}}$ and by `$\tilde{\nu}_{\overline{5}}$ resonance' we henceforth mean a
combination of both, unless stated otherwise. In this case, the $\tilde{\nu}_{\overline{5}}$(s) are
produced by tree-level interactions with valence quarks, as shown in Figure~\ref{figg12}. Subsequently, it may potentially
decay to $q \overline{q}$ with a large cross-section.  Such scenarios are severely constrained by
CMS dijet bounds~\cite{CMS:2015neg}. In addition, the charged scalar of the doublet will also contribute
to the dijet with an even larger cross-section at near the same invariant
mass~\cite{Allanach:2015ixl}.


Two such BPs are presented in the Table~\ref{tab:benchmarks2}. For these BPs, we assume that $\tilde{\nu}_{\overline{5}}$ couple
only to first generation valence quarks. The coupling $\eta^{D}_{115}$ relevant to $\tilde{\nu}_{\overline{5}}$ production (see Eq.~\ref{baryon_parity})
are chosen to be 0.07 and 0.08 for BP-4 and BP-5, respectively, by fixing $M^{E}_{10}=400$ GeV and $\lambda^{E}_{10}=1$ for the evaluation of the loop functions.
However, this results in an appreciable decrease in the width of the resonance. BP-4 represents a point wherein invisible decay of
$\tilde{\nu}_{\overline{5}}$ to SUSY particles is prohibited, and it provides a width of only 0.22 GeV. For completeness, we provide BP-5,
which corresponds to the largest width (0.35 GeV) possible for these kinds of scenarios when adopting the smallest $\sigma_{\gamma \gamma}$ that can fit the
data~\cite{Nilles}.
 We allow a small invisible
decay width of $\tilde{\nu}_{\overline{5}}$ to SUSY particles in order to achieve a larger width for BP-5.
We will discuss possible ways to resolve this problem in Sub-section~\ref{sneutrino}.
Nevertheless, a large invisible/semi-invisible decay width will further reduce an already small $\sigma_{\gamma \gamma}$ for $\tilde{\nu}_{\overline{5}}$
resonance scenarios and might not be able to fit the data. Clearly both these BPs are consistent with the 8 TeV CMS diphoton constraint~\cite{Nilles}.
In Table~\ref{dijet8v13} we also show the 8 TeV dijet cross-sections arising from these BPs, which is again within the CMS constraint
(1 pb for 750 GeV $q\overline{q}$ resonance). For dijet cross-section calculations in both Table~\ref{tab:benchmarks2} and Table~\ref{dijet8v13}
the contribution from the charged scalars of the doublet are also included.

Finally, we should mention that for all the calculations pertaining to $\tilde{\nu}_{\overline{5}}$ resonances are performed using
the Lagrangian of Eq.~\ref{baryon_parity}.  If we instead use the Lagrangian of Eq.~\ref{two_doublets}, a possible interaction of
$\tilde{\nu}_{\overline{5}}$ with $u \overline{u}$ is possible. However, due to strong constraints arising from dijet final
states, it will not alter qualitative discussions regarding the $\tilde{\nu}_{\overline{5}}$ resonance.

Coming back to the SM singlet scalar resonance, we notice from Table~\ref{tab:benchmarks} that the invisible decay,
$S \rightarrow\overline{N}N$, consumes the major fraction of the total width for BP-3.
This channel can be associated with a monojet process, $pp\rightarrow Sj \rightarrow j$+\met, where an extra jet from initial
state radiation, or gluon-splitting, can provide a large transverse momentum and boost the invisibly decayed $S$
into missing transverse energy. The monojet cross-section can be written as,
\be
\sigma_{\overline{N}Nj}(p_T) =\epsilon_{p_T}\times\left(\sigma_{\gamma\gamma}^{\text{obs.}} \cdot\frac{\text{BR}_{\overline{N}N}}{\text{BR}_{\gamma\gamma}}\right) \,,
\label{eq:monojet_xsec}
\ee
where $\epsilon_{p_T}$ is the cross-section ratio between $pp\rightarrow Sj$ with jet transverse momentum $p_T$ harder than a given threshold to $pp \rightarrow S$.
\be
\sigma_{\overline{N}Nj}(p_T) \equiv \sigma_{S}\times \mbox{BR}_{\overline{N}N}\times \epsilon_{p_T}
\ee
We obtain $\epsilon_{p_T}$ for various thresholds, as given in Table~\ref{tab:epsilon}. For monojet events, this jet $p_T$
equals the missing transverse energy \met.

\begin{table}[h]
\begin{center}
\setlength\tabcolsep{6pt}
\begin{tabular}{|c|c|c|c|}
\hline\hline
\met cut (GeV)& $\epsilon_{p_T}$ (8 TeV) &$\epsilon_{p_T}$ (13 TeV) & CMS~\cite{Khachatryan:2014rra} 8 TeV bound at 95\% C.L.\\
\hline
200&0.14 &0.18  & \\
300&0.063 &0.094 & \ \ 0.09 pb \\
400&0.031 &0.052 & \\
500&0.015 &0.030 & \  0.006 pb \\
\hline\hline
\end{tabular}
\caption{Parton level $\epsilon_{p_T}$  for monojet events with resonance at 750 GeV. MadGraph/MadEvent~\cite{Alwall:2014hca}
is used to simulate the monojet events. The production cross-section $\sigma_S$ is a factor of 5 smaller at the 8 TeV run.}
\label{tab:epsilon}
\end{center}
\end{table}
It is also interesting to note that the monojet cross-section falls faster than the CMS constraint, and a higher \met
cut gives a better constraint. Taking the upper limit with \met$>500$ GeV, the invisible decay branching is constrained to be,
\be
\frac{\text{BR}_{\gamma\gamma}}{\text{BR}_{\overline{N}N}} > \eta^{-1}\cdot \frac{\epsilon_{p_T}~ \sigma_{\gamma\gamma}}{0.006 \text{pb}} \sim 10^{-3} \text{ @ } 95\% \text{C.L,}
\label{eq:invisible_bound}
\ee
This constraint may be in tension with a large invisible width, which the measured $6\% M_S$ often requires.
In the next sub-section, we discuss options to evade this monojet constraint.

\subsection{Semi-invisible $S$ decays}





As a large invisible width in the $S$ decay may be constrained by monojet limits, it can be worth promoting
such invisible $\overline{N}N$ final state into `semi-'invisible, by allowing $N$ to decay into another
missing particle and relatively soft leptons arising dominantly from  $Z^\ast$ decay with $10-20$ GeV energy due to
small mass gaps between $N$ and the missing particle. The leptons can also be due to a slepton in between the NLSP and
LSP. Because the monojet searches veto on isolated leptons (e, $\mu$) with a small $p_T$ ($>$ 7 GeV at
CMS~\cite{Khachatryan:2014rra} and $>$ 10 GeV at ATLAS~\cite{ATLAS:2012zim}) and $p_T(\tau_h)>20$
GeV, the presence of such final state particles may allow semi-visible decays of this variety to
evade the monojet bounds.  Direct production (without monojet) of this event topology
from the resonance channel produces $N$'s back-to-back, resulting in a visible system consisting
solely of soft leptons, and an absence of appreciable missing energy.

Given the MSSM framework of our benchmark scenarios, the supersymmetric partner of $S$ may itself very ably serve
this purpose. For example, a $\kappa_S S^3$ type of term in the superpotential, as in the popular Next-to
MSSM (NMSSM)~\cite{bib:nmssm} model, allows the singlet to decay into a pair of singlinos, and the singlino
can mix with other gauginos and Higgsinos in the model.  As the singlino may derive its mass separately
when the singlet develops its own vacuum expectation value, the singlino can have a mass splitting of around $10-30$
GeV with Higgsinos and bino, with the latter being the dominant component of the LSP, while the
singlino dominates the NLSP. Also, one may imagine a slepton in between the NLSP and the LSP,
$\tilde\chi^0_2\rightarrow\tilde\chi^0_1 ll$ where the lepton energies depend on the mass gap between slepton and the $\tilde\chi^0$s.

In such a singlino-gaugino-Higgsino mixed case, a large self-coupling $\kappa_S\sim 1$ can dominate the $S$ width by
decaying into a pair of singlino-dominated NLSPs, which in turn produce high enough $p_T$ leptons to evade
monojet searches. Note that, beside the cubic $\kappa S^3$ term, the $\lambda_S S H_u H_d$ term also allows $S$
decay into neutrinos via their mixing with MSSM Higgses.
However, this interaction can induce $S\rightarrow VV,hh$ at branching
ratios comparable to that into neutrinos, which are highly constrained
by four lepton/$b$-jet searches.


\subsection{The $\tilde{\nu}_{\overline{5}}$ resonance case}
\label{sneutrino}

As mentioned earlier, the $\tilde{\nu}_5$ resonance scenarios discussed in Section~\ref{doublet} also suffer from the narrow width problem.
This problem can be resolved if one has near-degenerate sneutrinos with mass splitting of $\mathcal{O} (10)$ GeV.
Due to the antisymmetric nature of the $10\bar{5}\bar{5}$ couplings, either $\tilde{\nu}_{\overline{5}}$ mixes with MSSM sneutrinos
(for the superpotential of Eq.~\ref{baryon_parity}) or we require at least two $L_5$ doublets (for the superpotential of Eq.~\ref{two_doublets}).
Hence, it is natural to have two highly proximal scalar resonances. If this is the correct interpretation of the observed bump,
then with additional data it will resolve into two isolated narrow resonances.  We investigate the relative statistical evidence for such a scenario in the next section.

\subsection{Statistical analysis of resonance width and multiplicity}
\label{sec:dual}
In this section we consider the possibility of a wide resonance, and the
faking of a large apparent width by the overlap of adjacent dual resonances of more narrow intrinsic width.
For concreteness, we perform the analysis in the context of data collected by the
ATLAS~\cite{bib:ATLAS_diphoton} collaboration at the 13 TeV LHC,
representing an integrated luminosity of 3.2 ${\rm fb}^{-1}$.
Based upon this data, ATLAS reported a local $3.9\sigma$ excess at a diphoton invariant mass around
747~GeV, when assuming a large width of about 45~GeV (or $3.6\sigma$ when assuming a narrow width).
Whereas there is slight preference in the ATLAS analysis for a large decay width, the theoretical realization of
such a value presents certain challenges for maintaining consistency with limits in other channels,
as addressed in the present work.
We therefore presently investigate the goodness-of-fit achievable in a dual-resonance model, and quantify
the statistical preference for various mass gaps $\Delta M$ between the two peaks and various intrinsic widths $\Gamma$ for each.
We reiterate that data from CMS is not as suggestive of a large width,
although the new procedure that we introduce remains applicable
to any similar or future suggestion of an apparently large collider resonance width.

Table~\ref{tab:data_ATLAS} summarizes observations (extracted from the published plot) in the diphoton channel for eight
invariant mass $M_i^{\gamma\gamma}$ bins ($i = 1\ldots8$) of 40 GeV width spanning from 590 to 910 GeV,
which contain and surround the described local excess.  The binned signal counts $S_i$
represent deviation of the data from the estimated background $B_i$.
Also tabulated are an bin-by-bin estimates of the statistical uncertainty
$\sigma_i \equiv \sqrt{1+B_i}$ in the background projection
(taking ``+1'' as a regulator for faint backgrounds),
the signal significance $\aleph_i \equiv S_i/\sigma_i$,
and the fractional event count $f_i \equiv S_i/(S\equiv \sum_i S_i)$. 

The largest excess exists in the central pair of considered bins, whereas a less significant positive deviation is
observed in the two counting regions to either side.  Outside of this central region of interest
small (and frequently negative) deviations are observed.  We have tabulated an additional
four such bins (and have retained in the analysis a larger total of 17),
interpreted as a sideband, for the purpose of allowing overflow in scenarios
with especially wide Breit-Wigner distribution profiles and/or separation
between peaks of dual resonances.  In order to avoid propagating (ostensibly)
statistical fluctuations into quantities such as the mean resonance mass, the value of $S_i$
in these outer bins is forced to (0).  The inclusion or exclusion of additional
sideband bins has no meaningful effect on our analysis beyond the stated purpose of absorbing overflow.
We will adopt the hypotheses of a total beyond the SM event count $S = 14.8$, and a mean (significance weighted)
resonance mass $M \equiv \sum_i \VEV{M_i} \times \aleph_i \div \sum_i \aleph_i = 744$~GeV.
The unified local event significance is na\"{\i}vely estimated as $\aleph \equiv \surd\sum_i \aleph_i^2 \simeq 3.4$.

\begin{table}[h]
\bgroup
\setlength\tabcolsep{6pt}
{\footnotesize
\begin{tabular}{|c||c|c|c|c|c|c|c|c|}
\hline
$M_i^{\gamma\gamma}$~[GeV] & 590-630 & 630-670 & 670-710 & 710-750 & 750-790 & 790-830 & 830-870 & 870-910 \\ \hline
Background $B_i$ & 14.2 & 10.7 & 8.1 & 6.2 & 4.9 & 3.8 & 3.1 & 2.5 \\ \hline
Signal $S_i$ & 0.7 (0) & -0.7 (0) & 1.9 & 7.7 & 4.1 & 1.1 & -1.1 (0) & - 0.5 (0) \\ \hline
$\sigma_i \equiv \sqrt{1+B_i}$ & 3.9 & 3.4 & 3.0 & 2.7 & 2.4 & 2.2 & 2.0 & 1.9 \\ \hline
$\aleph_i \equiv S_i/\sigma_i$ & 0 & 0 & 0.63 & 2.9 & 1.7 & 0.50 & 0 & 0 \\ \hline
$f_i \equiv S_i/S$ & 0 & 0 & 0.13 & 0.52 & 0.28 & 0.07 & 0 & 0 \\ \hline
\end{tabular}
}
\egroup
\caption{
Summary of data collected by the ATLAS~\cite{bib:ATLAS_diphoton} collaboration at the 13 TeV LHC
in the diphoton channel with integrated luminosity of 3.2 ${\rm fb}^{-1}$.
$B_i$ represents the collaboration estimate for the SM background, and $S_i$ represents observed
deviations from this estimate.  }
\label{tab:data_ATLAS}
\end{table}

The normalized Breit-Wigner distribution function
$BW(E, \Gamma, M)$ in energy $E$ is written as follows. 
\begin{equation}
BW(E \Rightarrow \alpha M, \Gamma \Rightarrow \beta M, M) \equiv
\frac{2\beta}{\pi M}\times
\sqrt{\frac{2\,(1+\beta^2)}{1+\sqrt{1+\beta^2}}}
\times \frac{ 1 }{(\alpha^2-1)^2+\beta^2}
\label{eq:bw}
\end{equation}
Any given physical event detection environment will additionally be limited by constraints on resolving power,
which introduce stochastic fluctuation about the true signal shape.  This smearing effect may be modeled by the normalized
Gaussian distribution $G(x,\mu,\sigma)$ at $x$ with mean $\mu$ and standard deviation $\sigma$, as follows.
\begin{equation}
G(x,\mu,\sigma) \equiv \frac{1}{\sigma\sqrt{2\pi}}\times \exp\left\{-\frac{(x-\mu)^2}{2\sigma^2}\right\}
\label{eq:gauss}
\end{equation}

\begin{figure}[thp]
\centering \includegraphics[angle=0, width=14cm]{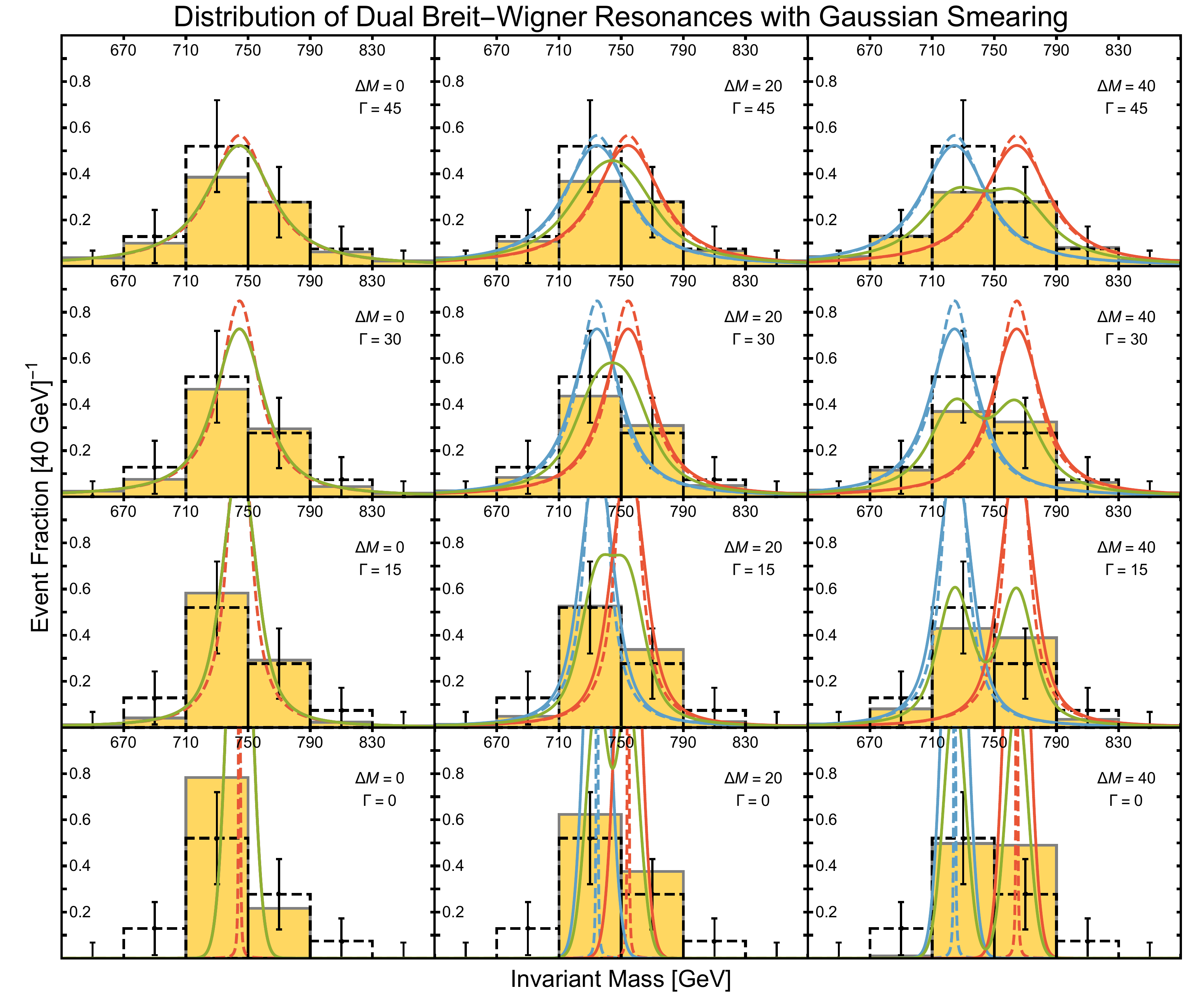}
\vspace{-0.3cm} \caption{
Event distributions are plotted for dual resonance scenarios with
various mass gaps $\Delta M$ and intrinsic widths $\Gamma$ for each resonance.
A mean resonance mass of $M = 744$~GeV is assumed in each case.
The blue and red curves depict the projected Breit-Wigner event yield
fractions per 40~GeV width before (dotted) and after (solid) convolving with
a 7~GeV Gaussian event resolution.  The solid green curve represents the mean
event profile for each scenario.  The gold-filled histograms represent projections
for the binned event counts.  The dotted black outlines represent experimental
excesses over the SM background, as observed by the ATLAS collaboration.
Error bars represent statistical fluctuations in the excess for each channel,
with a net observation of $S \sim 15$ events.
}
\label{fig:BWGrid}
\end{figure}

For analysis of the ATLAS diphoton data, we will adopt an intrinsic width of $\sigma = 7$~GeV. 
In order to predict the detector yield for a given theoretical distribution, it is necessary to convolve
with the smearing function, integrating over the original source spectrum.  We will denote the Gauss-convolved
Breit-Wigner distribution as $\widetilde{BW}(E, \Gamma, M, \sigma)$.
\begin{equation}
\widetilde{BW}(E, \Gamma, M, \sigma) \,\equiv\,
\int_0^\infty G(E,E',\sigma) \times BW(E', \Gamma, M) \, dE'
\label{eq:bwgauss}
\end{equation}
Integration of this distribution yields the event fraction $f_i^{BW}$ falling
within the $i^{th}$ energy bin interval $\left[E_i^{\rm Min},E_i^{\rm Max}\right]$.
\begin{equation}
f_i^{BW} = \int_{E_i^{\rm Min}}^{E_i^{\rm Max}} \widetilde{BW}(E, \Gamma, M, \sigma)\, dE
\label{eq:efrac}
\end{equation}
In order to incorporate the effects of dual similar resonances with separation $\Delta M$ about
a mean value $M$, it is simply necessary to replace
the smeared resonance function in Eq.~(\ref{eq:bwgauss}) with an average over two similar
such terms $\widetilde{BW}(E, \Gamma, M\pm\Delta M/2, \sigma)$, each displaced by half of the mass gap.
In Figure.~\ref{fig:BWGrid}, we exhibit the distinct intrinsic Breit-Wigner profiles,
as well as the effects of smearing and summing, and the projected binned event populations
for various values of $\Delta M$ and $\Gamma$.

\begin{figure}[thp]
\centering \includegraphics[angle=0, width=12cm]{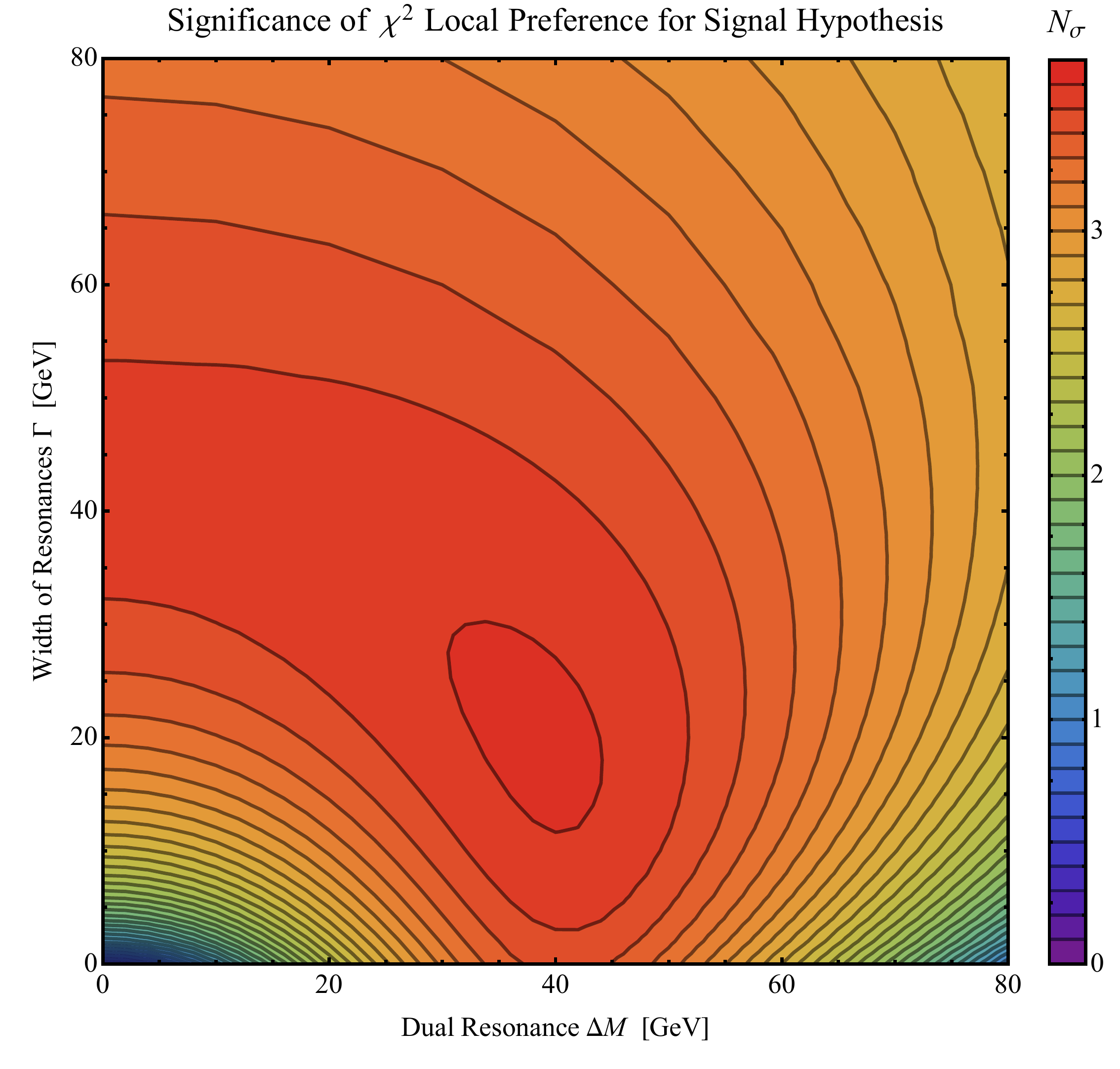}
\vspace{-0.3cm} \caption{
The local $\chi^2$ preference for each signal hypothesis over the null background-only hypothesis
is plotted for dual resonance scenarios with
various mass gaps $\Delta M$ and intrinsic widths $\Gamma$ for each resonance.
A mean resonance mass of $M = 744$~GeV is assumed in each case.
Significances for the log-likelihood difference in four degrees of freedom
are recast in terms of a multiple $N_\sigma$ of the one-dimensional standard deviation $\sigma$
corresponding to an equivalent $p$-value.
}
\label{fig:chisqsig}
\end{figure}

However, there is an essential element of fortune that effects the interpretation of binned event data.
Specifically, an intrinsically narrow resonance, broadened slightly by detector resolution effects,
and falling close to a binning edge, may perchance uniformly populate two adjacent bins and thereby
present the illusion of a larger width.  We may attempt to quantify this effect by marginalizing over a displacement
offset $\mathcal{O}$ of each bin by up to half of the interval $\Delta_E = E_i^{\rm Max} - E_i^{\rm Min}$
in either direction.  It is not sufficient, though, to simply take the mean event fraction gathered by 
the $i^{th}$ bin by averaging Eq.~\ref{eq:efrac} over all offsets $\mathcal{O}$.  This is because over and under population
of a bin will tend to cancel, whereas either scenario is individually to be disfavored.  Instead, it is
important to square prior to performing such an average, in order to effectively characterize the statistical
likelihood that the appearance of width in a binned analysis may be faked by a single narrow resonance near the boundary.

To proceed, we construct a $\chi^2$ statistic comparing the difference-squared of the observed signal $S_i$ to the
modeled event count $S_i^{BW}(\mathcal{O}) = S\times f_i^{BW}(\mathcal{O})$
at a given mutual offset $\left[E_i^{\rm Min}+\mathcal{O},E_i^{\rm Max}+\mathcal{O}\right]$ of the energy
boundaries of each bin
to the statistical variance $\sigma_i^2$, and sum over $C$ binned energy channels.
\begin{equation}
\chi^2(\mathcal{O}) \,\equiv\, \sum_{i=1}^C \frac{\left[S_i^{BW}(\mathcal{O})-S_i\right]^2}{\sigma_i^2}
\label{eq:chisq}
\end{equation}
We then marginalize this statistic over the binning offset,
holding the expected value $S_i$ of binned events, the statistical uncertainty $\sigma_i$,
and the resonance center $M$ fixed.
In practice, for computational speed, we opt to approximate this numerical integration as a
discretized sum with a reduced binning width (much less than the smearing width) of 1~GeV.
In order to avoid numerical divergences, a hard floor of $\Gamma \ge 0.05$~GeV is imposed for all calculations.
\begin{equation}
\chi^2 = \frac{1}{\Delta E} \times \int_{-\Delta E/2}^{+\Delta E/2} \chi^2(\mathcal{O})\,d\mathcal{O} 
\label{eq:emargin}
\end{equation}

This statistic is dependent upon four degrees of freedom, namely the 
mass gap $\Delta M$ between the dual resonances, the intrinsic width $\Gamma$ of
each resonance, as well as the mean resonance mass (which we take to be optimized at $M = 744$~GeV)
and the net signal strength (which we take to be optimized at $S = 14.8$ events).
For the null background-only model, which has no degrees of freedom, Eq.~(\ref{eq:chisq})
reduces to the previously tabulated $\aleph^2 \simeq 11.7$.
The goodness of a fit to data may be quantified by the comparison of these factors.
Specifically, Wilks' theorem states that twice the difference of negative-log-likelihoods
for nestable model templates is approximately $\chi^2_D$ distributed,
with degrees of freedom $D$ equal to the difference in number of optimized parameters.
The criterion for a significant improvement at a type-I error level $p$
for the 4-parameter signal fit over the background-only template follows, where
the cumulative distribution function (CDF) gives the fraction of parameter space bounded
within a multi-dimensional (typically Gaussian) integration out to some ``radius'' $\chi$.
\begin{equation}
\chi^2
\,\le\,
\aleph^2 -
{\rm CDF}^{-1}\big(D\Rightarrow 4,1-p\big)/2
\label{eq:wilks}
\end{equation}
The inverse CDF is simply the $\chi^2_D$ boundary value in $D$ dimensions
for which a fraction $p$ of possible outcomes would be considered more extreme.
For example, with $D=4$, the inverse CDF $\chi^2$-values for confidences levels $(1-p)$
corresponding to $\{68,95,99.7\}\%$, i.e. $\{1,2,3\}$ standard deviations, are $\{ 4.7,9.7,16.3 2\}$, respectively.
Eq.~(\ref{eq:wilks}) may alternatively be inverted to solve for the $p$-value,
which may be converted into an equivalent significance multiple $N_\sigma$
of the one-dimensional Gaussian standard deviation $\sigma$.
\begin{align}
p \,&=\, 1 - {\rm CDF}\big(\,D\Rightarrow 4,\,
\label{eq:wilksinv1}
2\times\big[\aleph^2 - \chi^2\big]\,\big) \\
N_\sigma \,&=\, \sqrt{\,{\rm CDF}^{-1}\big(D\Rightarrow 1,1-p\big)\,}
\label{eq:wilksinv2}
\end{align}

Contours of the value of $N_\sigma$ from Eq.~(\ref{eq:wilksinv2}) are plotted in Figure~\ref{fig:chisqsig}
for relevant points in the two-dimensional ($\Delta M$,$\Gamma$) subspace, at fixed $M$.
The best fit is found to occur for a dual resonance with $\Delta M \simeq 40$~GeV,
and an intrinsic width of each resonance $\Gamma \simeq 20$~GeV.  All scenarios with
$\Gamma \gtrsim 20$~GeV fit the data reasonably well.  With $\Gamma \lesssim 10$~GeV
a good fit can apparently not be realized for the single resonance model. 
With a single resonance, $\Gamma \simeq 40$~GeV is preferred, and with two narrow
resonances, $\Delta M \simeq 40$~GeV is preferred.
In general, there is a band of favored solutions, wherein the introduction
of a modest splitting between dual resonances allows for the intrinsic width of each to be reduced.

\section{Conclusion}

We have considered vector-like
multiplets $(5,\overline{5})$ and $(10,\overline{10})$
in the context of $SU(5)$ gauge coupling unification and investigated their compatibility with the 750 GeV diphoton
resonance, using the renormalized masses and Yukawa couplings at that scale.
\begin{itemize}
\item  We demonstrated the effect of these new multiplets on the unified scale and coupling strength.
We also investigated the new Yukawa couplings and mass terms associated with new vector-like
multiplets and the new scalar perturbatively, evolving down universal values from the GUT scale.
\item Inclusion of the new vector-like multiplets allows us to reduce the burden on the stop squarks to provide
additional necessary contribution to the 125 GeV Higgs mass in the MSSM. We have provided a numerical estimate of
contributions to the Higgs mass arising from various vector-like configurations.
\item We also showed that the proton decay rate for $p\rightarrow e^{+}\pi^0$ in these models may
be enhanced and lie within the reach of future proton decay experiments.
\item We demonstrated the capacity of 3 copies of $(5,\overline{5})$ and
1 copy of $(10,\overline{10})$ to explain the observed excess.
However, the width associated with such a resonance is very
narrow $\lesssim 1$ GeV for both $3 \times (5,\overline{5})$ and $(10,\overline{10})$,  whereas
the experimentally preferred width is much larger.
\item In order to accommodate such a width, we introduced an additional decay mode where the new scalar singlet decays into
singlinos, which then decay into Higgsinos by emitting soft leptons with $p_T \sim 10-20$ GeV.
This scenario is not excluded by monojet or other constraints.
\item In addition, we also showed that using components from different multiplets of $(10,\overline{10})$ and
$(5,\overline{5})$ (without gauge coupling unification), we can write down a new interaction where the
neutral component of the new lepton doublet scalar is responsible for the resonance. In such a
scenario, R-parity is preserved. We showed that this model explains both 8 and 13 TeV diphoton excess data for a 750 GeV resonance. The diphoton cross-section in this scenario is bounded from above by the dijet constraint to $\sim 1$ fb.
The decay width can be enhanced by 
having two new $L_5$ doublets or mixing between one $L_5$ doublet and MSSM doublets $L_i$, due to the antisymmetric nature of the coupling $10\bar{5}\bar{5}$. In such a case, we may naturally realize two adjacent resonances, and no additional contribution to the width is required.
\item We presented a new statistical approach to quantifying the relative preference for various decay width
scenarios and mass gaps in the context of scattering through a dual resonance.  Specifically, this method
is designed to cope with the difficulties presented by narrow resonances interacting with energy bin boundaries. 
\end{itemize}

Subsequent to the analysis performed in the current work, it was simultaneously
announced by the ATLAS~\cite{ICHEP_ATLAS} and CMS~\cite{ICHEP_CMS} collaborations
that the diphoton excess observed around $750$~GeV during the 2015 run at LHC had not
carried over into the 2016 data. 
The current status at ATLAS is 2.4~$\sigma$ (local), when combining 2015+2016 datasets, reduced
from 3.9~$\sigma$ during the 2015 run alone~\cite{ICHEP_ATLAS}, and the current status at CMS
is 1.9~$\sigma$ (local), when combining 8~TeV+2015+2016 datasets, reduced from 3.4~$\sigma$
during the 8~TeV+2015 runs alone~\cite{ICHEP_CMS}.  This is
indicative of just how profoundly suggestive
the 2015 diphoton data was, such that its influence persists even when integrated
with the four-fold larger null result from 2016.
The compelling coincidence of data from both experiments (largely mitigating look-elsewhere considerations), and
the rather unexpected features of the data (including the pure photon final state, the implied cross section,
the evasion of competing bounds from monojet and related searches, the introduction of a new scale,
and the suggestion of a rather anomalously large width) presented a fascinating phenomenological
puzzle, which inspired a wide range of creative analyses in pursuit of a consistent and naturally
motivated framework capable of matching the observations.  Many of the ideas introduced during this
period of intense study and dialog within the high energy physics community will certainly outlive in relevance
the passing of the particular circumstances that originally inspired them.  As just one specific
example (for which we do not advocate in particular), ATLAS now observes a 2.4~$\sigma$ local
excess in the diphoton channel around 1.6~TeV, when combining the 2015+2016 data.  If some such similar excess at
larger mass (or smaller cross-section) were to be persuasively established by the future collection
of additional data, then self-consistent ideas proposed in association with the 750~GeV diphoton
could take on new relevancy at the new physics scale.


\begin{acknowledgments}
We thank Teruki Kamon and Keith Ulmer for helpful discussions. This work is supported in part by DOE grant numbers
DE-FG02-13ER42020 (B.D., T.G.) and DOE-SC0013880 (Q.S.),
Bartol Research Institute (I.G.), the  Rustaveli National Science Foundation  No. 03/79 (I.G.),
Natural Science Foundation of China grant numbers 11135003, 11275246, and 11475238 (T.L),
National Science Foundation grant number PHY-1521105 (J.W.W.),
and the Mitchell Institute for Fundamental Physics and Astronomy (Y.G.).

\end{acknowledgments}

\end{document}